\documentclass[preprint,preprintnumbers,amsmath,amssymb,floatfix,nofootinbib]{revtex4}
\usepackage{epsf}
\usepackage{graphicx}
\usepackage{subfigure}
\usepackage{bbold}

\newcommand{\bea}{\begin{eqnarray}}
\newcommand{\eea}{\end{eqnarray}}
\newcommand{\nn}{\nonumber \\}
\newcommand{\lag}{\ensuremath{{\cal L}}}
\newcommand{\Tr}{\operatorname{Tr}}

\def\beq{\begin{equation}}
\def\eeq{\end{equation}}
\begin{document}

\title{Unravelling an extended quark sector through multiple Higgs production?}
\author{S.~Dawson}
\author{E.~Furlan}
\author{I.~Lewis}

\affiliation{
Department of Physics, Brookhaven National Laboratory, Upton, New York, 11973
\vspace*{.5in}}

\date{\today}

\begin{abstract}
In many new physics scenarios, the particle content of the Standard Model is extended and the Higgs 
couplings are modified, sometimes without affecting single Higgs production. We analyse two models 
with additional quarks.
In these models, we compute double Higgs production from gluon fusion 
exactly at leading-order, and present analytical results 
in the heavy-quark mass approximation. The experimental bounds from
precision electroweak measurements and from the measured rate of single Higgs production combine
to give significant restrictions for the allowed deviation of the  double Higgs production rate from the Standard 
Model prediction as well as on the branching ratio for the Higgs decay into photons. 
The two models analysed eventually present a similar Higgs phenomenology as the Standard Model. 
We connect this result to the magnitude of the dimension six operators contributing to the 
gluon-fusion Higgs production. 
\end{abstract}

\maketitle

\section{Introduction}
The search for the source of electroweak symmetry breaking has dominated particle theorist's efforts for 
decades. Now that a particle with many of the right properties to be the Higgs boson of the Standard 
Model has been discovered~\cite{:2012gk,:2012gu}, the efforts turn to understanding the properties of this particle. In the Standard
Model, the couplings of the Higgs boson to fermions, gauge bosons, and to itself are firm predictions of the
model. In models with new physics, however, these couplings can be different. 

The dominant production mechanism for a Higgs boson is gluon fusion, which
is sensitive to many types of new physics. The simplest possibility is for new heavy
colored scalars~\cite{Kribs:2012kz,Dobrescu:2011aa} and/or 
fermions~\cite{Dawson:2012di,Cacciapaglia:2011fx,Carmi:2012yp,Barger:2012hv,Azatov:2012rj,
Kumar:2012ww,Wang:2012gm,Bonne:2012im,Voloshin:2012tv,Bertuzzo:2012bt,Moreau:2012da}
to contribute to Higgs production.
However, since the observed Higgs candidate particle is produced at roughly the 
Standard Model rate, extensions of the Higgs sector
beyond the Standard Model are extremely constrained.
For example, a model with a sequential fourth generation of chiral fermions predicts large deviations in the Higgs 
rates ~\cite{,Anastasiou:2010bt,Anastasiou:2011qw,Djouadi:2012ae,Denner:2011vt,Eberhardt:2012gv} 
and is excluded by the limits
on Higgs production for any Higgs mass below around 600~GeV~\cite{ATLAS-CONF-2011-135,SM4-LHC}. 
The properties of these potential new colored particles
are further limited by precision electroweak measurements.
Models in which the Higgs boson is 
composite~\cite{Hill:1991at,Dobrescu:1997nm,Chivukula:1998wd,He:1999vp,He:2001fz,
Hill:2002ap,Agashe:2004rs,Agashe:2006at,
Giudice:2007fh,Espinosa:2010vn,Grober:2010yv,Gillioz:2012se},
along with models which generate
new higher dimension effective operators involving the Higgs boson and gluons~\cite{Manohar:2006gz,Pierce:2006dh}, 
can also induce a single Higgs production rate different from that of the Standard Model.
Untangling the source of possible deviations from the Standard Model by measuring the production
and decay rates of the Higgs boson will be quite difficult in models where there are only small
differences from the Standard Model predictions.  

In this paper, we examine
the extent to which the gluon fusion production of two Higgs bosons can have a rate very different from
that predicted by the Standard Model~\cite{Plehn:1996wb,Glover:1987nx}, given the 
restrictions from electroweak precision physics and
from single Higgs production. The observation of double Higgs production via gluon fusion is important
in order to measure the cubic self coupling of the Higgs boson~\cite{Djouadi:1999rca,Dolan:2012rv}. In the
Standard Model, the rate is small, although the ${\cal O}(\alpha_s^3)$ radiative corrections are known in the infinite
top quark mass limit and are large~\cite{Dawson:1998py,Binoth:2006ym}. 
For a 125~GeV Higgs particle, the most likely channel for $HH$
exploration is $gg\rightarrow HH\rightarrow b{\overline b} 
\gamma \gamma$~\cite{Baur:2003gp}, 
where studies have estimated
that the LHC at full energy will be sensitive to this process
with around 600~fb$^{-1}$.
Using jet substructure techniques, the 
$HH\rightarrow b {\overline b} W^+W^-$ and $HH\rightarrow b {\overline b} \tau^+\tau^-$ channels may be available 
with about 600~fb$^{-1}$\cite{Papaefstathiou:2012qe} and 1000~fb$^{-1}$\cite{Dolan:2012rv}. 
This is clearly not physics which will be
done during the early phase of LHC operations, unless the rate is significantly larger than in the Standard 
Model~\cite{Contino:2012xk}.

Double Higgs production can further be studied through vector boson fusion, which is also sensitive to the three Higgs
self coupling~\cite{Contino:2010mh}. Vector boson fusion production of two Higgs bosons can be affected by new operators 
involving the $W$ and $Z$
gauge bosons and the Higgs, but is not sensitive to the new colored particles which contribute to the gluon fusion process. 
Hence the two production mechanisms can provide complementary information.

Double Higgs production from gluon fusion first occurs at one loop and is therefore
potentially modified by the
same new heavy
colored particles which contribute to single Higgs production. However, as pointed out in Ref.~\cite{Pierce:2006dh},
single and double Higgs production are sensitive to different higher dimension effective operators and in principle, 
the single Higgs production rate could be Standard Model like, while the double Higgs production could be
highly suppressed or enhanced. Here, we consider the effects of both heavy vector-like and chiral colored fermions
on the single and double Higgs production rates, and the interplay between them. We will not consider models with 
extended Higgs sectors,
or with higher dimension non-renormalizable operators.

For single Higgs production, it is useful to analyze the effects of non-Standard Model colored particles using
a low energy theorem~\cite{Kniehl:1995tn}. 
The theorem can be formulated using the background field method in terms of the traces of
the mass matrices of colored objects, which eliminates the need
to diagonalize complicated mass matrices~\cite{Low:2009di}. 
The low energy theorem can be extended to double Higgs production, where new
features arise~\cite{Gillioz:2012se}. 
In models with extended fermion 
sectors (for example, in little Higgs models~\cite{ArkaniHamed:2002qy,Low:2002ws,Perelstein:2003wd,
Chang:2003un,Chen:2003fm,Hubisz:2005tx,Han:2005ru})
 there are contributions to double Higgs production containing
more than one flavor of fermion~\cite{Dib:2005re}. These diagrams contain
axial couplings to the Higgs boson which are non-diagonal in 
the fermion
states and we demonstrate how these effects can be included using a low energy theorem. 
Low energy theorems are extremely useful for single Higgs production and generally give estimates of the total cross
section which are quite accurate. For double Higgs
production, however, the low energy theorems provide an estimate of the total rate which typically disagrees with the exact rate by
$50\%$ or more. The low energy theorem does not reproduce kinematic distributions accurately, but instead predicts 
high energy tails which are not present in the full theory~\cite{Baur:2002rb}.

In this paper, we study the effects of heavy colored fermions on the gluon fusion double Higgs production rate and
show that agreement with single Higgs production requires the double Higgs rate to be close to
that of the Standard Model. We demonstrate how this can be understood in terms of the effective operator approach of 
Ref.~\cite{Pierce:2006dh} and discuss the limitations of the low energy theorem for $gg\rightarrow HH$.
Interestingly, composite Higgs models and little Higgs models
receive potentially large corrections to the $gg\rightarrow HH$ process
 from the non-renormalizable operator $t{\overline {t}} HH$. The observation of such a large effect would be a ``smoking gun''
 signal for such models~\cite{Grober:2010yv,Contino:2012xk,Gillioz:2012se}.

\section{Double Higgs production}
\subsubsection{The Standard Model}
\label{sec21_gghh_SM}
\begin{figure}[tb]
\subfigure{
      \includegraphics[width=0.3\textwidth,clip]{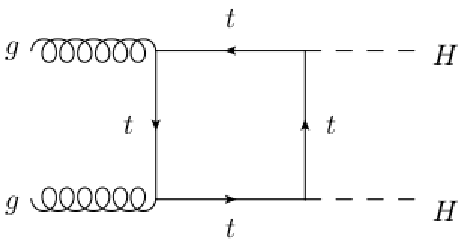}
}
\subfigure{
      \includegraphics[width=0.3\textwidth,clip]{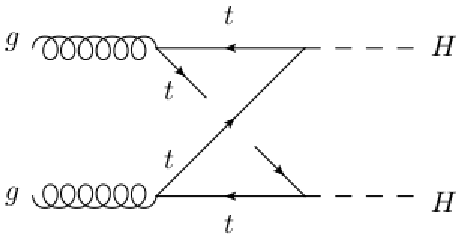}
}
\subfigure{
      \includegraphics[width=0.3\textwidth,clip]{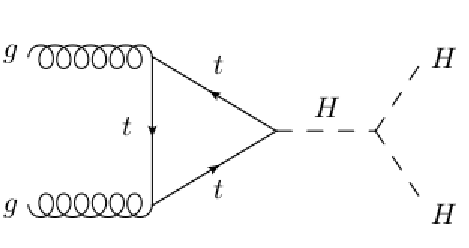}
}\caption{Feynman diagrams for $gg \to HH$ in the Standard Model.}
\label{diags:gghh_topologies}
\end{figure}
In the Standard Model, double Higgs production from a gluon-gluon initial state
arises from the Feynman diagrams shown in Fig.~\ref{diags:gghh_topologies}.
The result is sensitive to new colored objects (fermions or scalars) in the loops and to the 
Higgs trilinear self-coupling.
The amplitude for $g^{a,\mu}(p_1)g^{b,\nu}(p_2)\rightarrow H(p_3)H(p_4)$ is 
\begin{equation}
	A^{\mu\nu}_{ab}={\alpha_s\over 8 \pi v^2}\delta_{ab}\biggl[P_1^{\mu\nu}(p_1,p_2)F_1(s,t,u,m_t^2)+
	P_2^{\mu\nu}(p_1,p_2,p_3)F_2(s,t,u,m_t^2)\biggr]\; ,
\label{eq:amp}
\end{equation}
where $P_1$ and $P_2$ are the orthogonal  projectors onto the spin-$0$ and spin-$2$ states respectively,
\begin{eqnarray}
	P_1^{\mu\nu}(p_1,p_2)&=&
	g^{\mu\nu}-{p_1^\nu p_2^\mu\over p_1\cdot p_2} \; , \nn
	P_2^{\mu\nu}(p_1,p_2,p_3)&=&g^{\mu\nu}+{2\over s p_T^2 } \left(
		m_H^2 p_1^\nu p_2^\mu 
		- 2 p_1.p_3 \, p_2^\mu p_3^\nu - 2 p_2.p_3 \, p_1^\nu p_3^\mu + s \, p_3^\mu p_3^\nu
\right)\, ,
\end{eqnarray}
$s,t$, and $u$ are the partonic Mandelstam variables,
\beq
	s = (p_1+p_2)^2 \; , \quad
	t = (p_1-p_3)^2 \; , \quad
	u = (p_2-p_3)^2 \; ,
\eeq
$p_T$ is the transverse momentum of the Higgs particle,
\beq 
	p_T^2={ut-m_H^4\over s} \; ,
\eeq
and $v=(\sqrt{2}G_F)^{-1/2}=246$~GeV.
The functions $F_1$ and $F_2$ are known analytically~\cite{Glover:1987nx,Plehn:1996wb}. 
Finally, the partonic cross section is given by 
\begin{eqnarray}
	{d{\hat\sigma}(gg\rightarrow HH)\over dt}
	&=&
	{\alpha_s^2\over 2^{15}\pi^3v^4}
	{| F_1(s,t,u,m_t^2)|^2+| F_2(s,t,u,m_t^2) |^2\over s^2} \; ,
\end{eqnarray}
where we included the factor of ${1\over 2}$ for identical particles in the final
state.

In the Standard Model,  the chiral fermions are
\begin{equation}
	\psi^i_L=\left(\begin{matrix}
	u_L^i\\
	d_L^i\end{matrix}\right), \quad u^i_R, \,d^i_R\, ,
\end{equation}
where $i=1,2,3$ is a generation index and the Lagrangian describing the quark masses is 
\begin{equation}
	-{\cal L}_M^{SM}=\sum_{i}  \lambda^d_{i} {\overline{\psi}}^i_L \Phi d^i_R+
	\lambda^u_{i}{\overline{\psi}}^i_L {\tilde \Phi} u^i_R+ {\rm h.c.} \; .
\end{equation}
Here $\Phi = \left( \phi^{+}, \phi^0 \right)^T$ is the Higgs doublet, ${\tilde \Phi}=i\sigma_2 \Phi^*$ 
and $\phi^0 = {v+H\over\sqrt{2}}$.
Note that in the Standard Model the Higgs couplings $\lambda^{u,d}_{i}$ are purely scalar. 
In the following we will focus on the third generation quarks and use the standard notation 
$u^3 = t$, $d^3 = b$, with $\lambda^d_{3} \equiv \lambda_1$ and 
$\lambda^u_{3} \equiv \lambda_2$.

In the Standard Model, the dominant contributions come from top quark loops.
Analytic expansion of the amplitudes in the limit $m_t^2 >> s$  yields the leading
terms
\begin{eqnarray}
\label{eq:gghh_SM_expansion}
	F_1(s,t,u,m_t^2)&\equiv &F_1^{tri}(s,t,u,m_t^2)+F_1^{box}(s,t,u,m_t^2) \; ,\nn
	F_1^{tri}(s,t,u,m_t^2)&=& {4 m_H^2\over s - m_H^2} s
		\biggl\{1+{7\over 120}{s\over m_t^2}+{1\over 168}{s^2\over m_t^4}
		+{\cal O}\biggl({s^3\over m_t^6}\biggr)\biggr\} \; ,\nn
	F_1^{box} (s,t,u,m_t^2)&=&-{4\over 3}s\biggl\{ 1+{7\over 20}{m_H^2\over m_t^2}
					+{90 m_H^4-28 m_H^2 s+12 s^2-13p_T^2s \over 840 m_t^4}
					+{\cal O}\biggl({s^3\over m_t^6}\biggr)\biggr\} \; ;\nn
	F_2(s,t,u,m_t^2) & = & - {11\over 45} s {p_T^2\over m_t^2} \biggl\{ 
			1 + {62 m_H^2 - 5 s \over 154 m_t^2}
			 +{\cal O}\biggl({s^2\over m_t^4}\biggr)
	\biggr\} \; .
\label{fdef}
\end{eqnarray}
The leading terms in the inverse top mass expansion 
of  Eq.~\ref{fdef} are called the ``low energy theorem'' result and give the 
$m_t$-independent amplitudes~\cite{Glover:1987nx,Plehn:1996wb}
\begin{eqnarray}
F_1(s,t,u,m_t^2)\mid_{LET}
&\rightarrow& \biggl(-{4\over 3}+{4m_H^2\over s-m_H^2}\biggr)s \; ,
\nonumber \\
F_2(s,t,u,m_t^2)\mid_{LET}&\rightarrow & 0\; .
\end{eqnarray}

From Eq.~\ref{fdef}, we can clearly see that the triangle diagram has no angular dependence and only makes an 
$s$-wave contribution.  This result is expected since the triangle diagram has a triple-scalar coupling, which has 
no angular momentum dependence.  For the box diagrams, at the lowest order in $F_2^{box}$ there is angular 
momentum dependence reflected in $p_T^2$, which is expected from the spin-2 initial state and spin-0 final state.  
At ${\cal O}(m_t^{-4})$ in $F_1^{box}$ there is also an angular momentum dependent piece proportional to $p_T^2$.  
Since the initial and final states for the $F_1$ contribution are both spin-0, this is a somewhat surprising result.  To gain 
insight into the angular dependence of $F_1^{box}$ and further insight into $F_2^{box}$, the functions can be 
decomposed into Wigner $d$-functions, $d^j_{s_i,s_f}$, where $j$ is the total angular momentum and $s_i$ ($s_f$) is 
in the initial (final) state spin:
\begin{eqnarray}
\label{eq:gghh_SM_dfunc}
F_1^{box}(s,t,u,m_t^2)&=&-{4\over 3}s\left[\left(1+{7\over 20}{m_H^2\over m_t^2}+
{540 m_H^4-116 m_H^2 s+59 s^2\over 5040 m_t^4}\right)d^0_{0,0}(\theta) \right. \nn 
&&\qquad + \left. {13 s^2-52 m_H^2 s\over 5040 m_t^4}d^2_{0,0}(\theta)+{\cal O}\left({s^3\over m_t^6}\right)\right] \; ,\nn
F_2^{box}(s,t,u,m_t^2)&=&-{11\over 45}s {s-4 m_H^2\over \sqrt{6} m_t^2}\left[1+{62 m_H^2-5 s\over 154 m_t^2}
+{\cal O}\left(s^2\over m^4_t\right)\right] d^2_{2,0}(\theta) \; .
\end{eqnarray}
Here $\theta$ is the angle between an initial state gluon and final state Higgs, 
\beq
	t=m_H^2-\frac{s}{4} \left(1-\beta \cos \theta \right) 
	\quad {\rm and} \quad 
	\beta = \sqrt{1- \frac{4 m_H^2}{s}} \;.
\eeq
In $F_1^{box}$, we can see the expected spin-0 $s$-wave component, $d^0_{0,0}$, and an additional spin-0 $d$-wave 
component, $d^2_{0,0}$, at ${\cal O}(m_t^{-4})$.  The $s$-wave and $d$-wave components are orthogonal. 
Hence any angular independent observables, such as total cross section and invariant mass distribution, are independent of the 
$p_T^2$ component of $F_1^{box}$ up to ${\cal O}(m_t^{-8})$.  Finally, $F_2^{box}$ is wholly dependent on the initial 
state spin-2 $d$-wave function $d^2_{2,0}$, as expected from Eq.~\ref{eq:amp}.

\begin{figure}[tb]
\begin{center}
\includegraphics[scale=0.4]{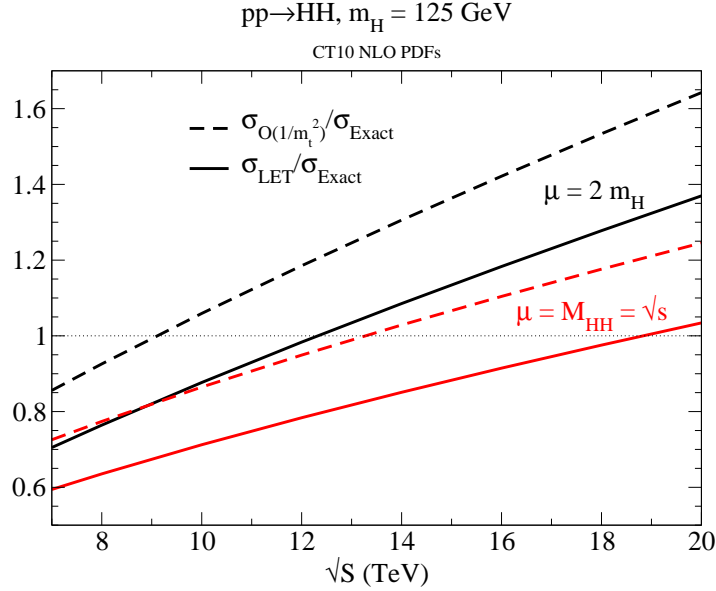}
\vskip .05in
\caption{Double Higgs production cross section as a function of the hadronic center of mass energy $\sqrt{S}$ in the 
	infinite top mass approximation, LET,  (solid lines) and retaining the ${\cal O} \big({s\over m_t^{2}}\big)$ corrections 
	(dashed lines), 
	normalized to the exact result.  The black (red) curves choose as the renormalization and factorization 
	scales $\mu=2m_H$ ($\mu=M_{HH}=\sqrt{s}$). }
\label{fg:sig_tot_sm}
\end{center}
\end{figure}

In Fig.~\ref{fg:sig_tot_sm}, we compare the total cross section for double Higgs production at different orders in the 
large mass expansion against the exact result\footnote{The exact result always includes the contributions from both 
the top and bottom quarks.}, as a function of the center of mass energy in $pp$ collisions. 
We use the CT10 NLO PDF set~\cite{Lai:2010vv} and run the strong coupling constant through NLO 
from its value $\alpha_s(m_Z ) = 0.118$. We fix $m_t =$~173~GeV and  $m_b =$~4.6~GeV.  
The low energy theorem results are quite sensitive to the scale choice, and typically 
reproduce the exact results to within  roughly $50\%$ error.
This ``agreement" between the infinite mass approximation (LET) and the exact result
is not improved by the inclusion of higher orders in the large mass expansion.
In single Higgs production, the reliability of the infinite mass approximation has been investigated through 
NNLO~\cite{Harlander:2009mq, Pak:2009dg,Harlander:2009my,Pak:2011hs}. Because of the shape of the gluon 
parton luminosity, which peaks at large values of $x = m_H^2/s$ and decreases rapidly, 
the largest contribution to the hadronic single Higgs cross section comes from the region below the top quark threshold, $s < 4 m_t^2$, where 
the large top mass approximation holds. As a consequence, finite mass corrections to single Higgs production have 
an effect of less than $1\%$.   On the other hand,  for double Higgs production the partonic energy is always $s>4m_H^2$ and
the condition for validity of the low energy theorem, $s\ll 4m_t^2$, is typically not satisfied.

\begin{figure}[tb]
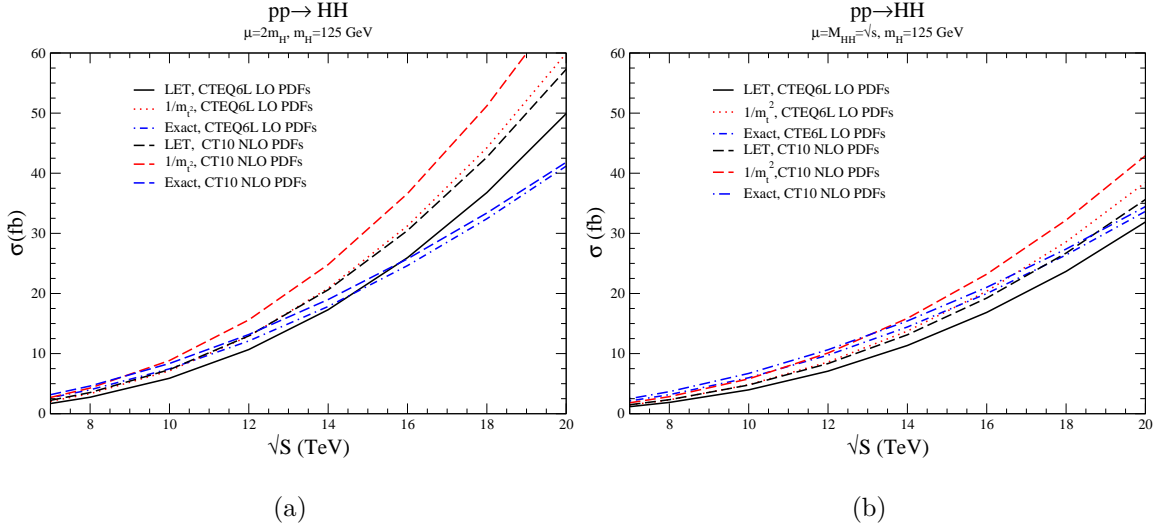

	\subfigure[]{\includegraphics[scale=0.45,clip]{sm_pdf.eps}}
	\subfigure[]{\includegraphics[scale=0.45,clip]{sm_pdf_rts.eps}}
	\caption{Total cross sections for $HH$ production using CTEQ6L LO PDFS and CT10 NLO PDFs.  The renormalization/factorization scale is 
	$\mu=2m_H$ in (a) and $\mu=M_{HH}=\sqrt{s}$ in (b). For all curves, $\alpha_s$ is evaluated at NLO. 
	}
	\label{fg:sm_hh_pdf}
\end{figure}
Fig.~\ref{fg:sm_hh_pdf} shows the sensitivity of the results to the choice of the PDF sets. The exact 
result has a small sensitivity to the choice of LO vs NLO PDFs.  However, the infinite mass limit (LET) 
of the result is quite sensitive to the choice of PDFs.
Including higher order terms in the top mass expansion does not reduce this sensitivity to the choice of PDFs.

\begin{figure}[tb]
\begin{center}
\includegraphics[scale=0.5]{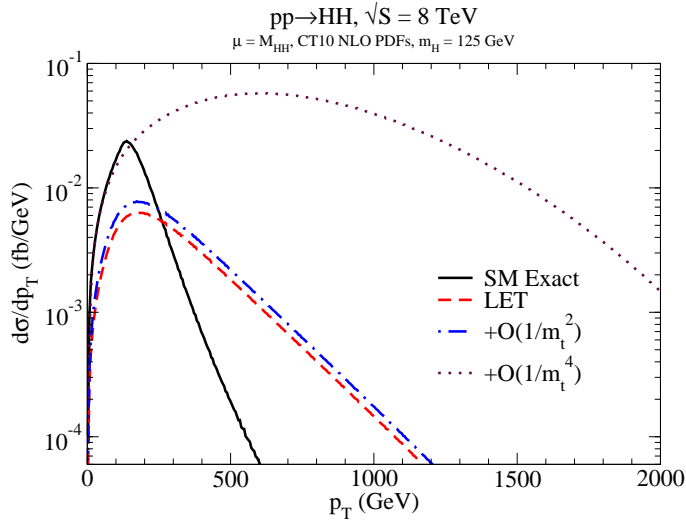}
\vskip .05in
\caption{Transverse momentum distribution for double Higgs production cross section.
	The Standard Model exact result, the LET and the heavy top mass approximations 
	up to ${\cal O} ( m_t^{-4})$ are shown.
	We choose as the renormalization and factorization 
	scales $\mu=M_{HH}=\sqrt{s} $  and use the CT10 NLO PDFs.}
	\label{fg:sig_pt_sm}
\end{center}
\end{figure}

The inadequacy of the infinite mass approximation for double Higgs production becomes even more apparent when 
looking at kinematic distributions~\cite{Baur:2002rb}.  Consider for example the invariant mass of the $HH$ system,  
\begin{eqnarray}
	{d\sigma(pp\rightarrow HH)\over dM_{HH}}
	&=&
	{2M_{HH}\over S}
	{\hat{\sigma}}(gg\rightarrow HH)
	{d{\cal L}_{gg}\over d\tau} 
\quad , \quad \nn
	{d {\cal L}_{gg}\over d\tau} &=&
	\int_\tau^1 {dx\over x}
	g(x,\mu_F)g\biggl({\tau\over x},\mu_F\biggr) \; ,
\end{eqnarray}
where $S$ is the hadronic center of mass energy squared, $M_{HH}=\sqrt{s}$, and
$\tau={s\over S}$.
\begin{figure}[b]
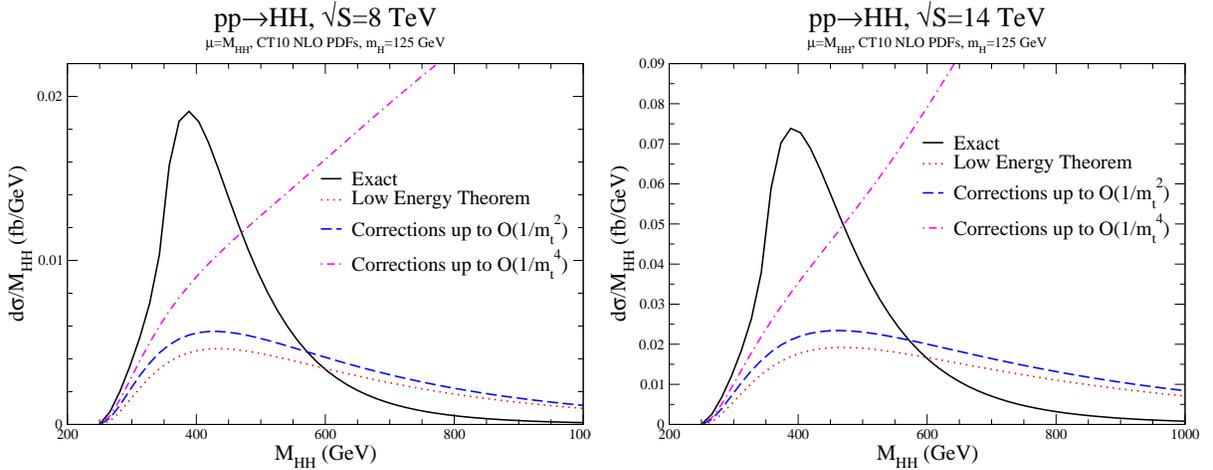

	\subfigure{\includegraphics[scale=0.45]{hh_SM_8.eps}}
	\subfigure{\includegraphics[scale=0.45]{hh_SM_14.eps}}
	\caption{Invariant mass distributions for Higgs pair-production at $\sqrt{S}=8$~TeV and $\sqrt{S}=14$~TeV, for terms in the 
		large mass expansion up to ${\cal O}(m_t^{-4})$ (Eq.~\ref{eq:gghh_SM_expansion}) and  with the full mass dependence.}
	\label{fg:sm_hh}
\end{figure}
In Fig.~\ref{fg:sm_hh} we analyse the impact of the finite mass corrections to the invariant mass distribution 
at the $\sqrt{S}=8$~TeV  and $\sqrt{S}=14$~TeV LHC. 
The inclusion of the ${\cal O}(m_t^{-2})$ corrections does not significantly improve the low energy 
 theorem results. The $m_t^{-4}$ terms
 fail entirely in reproducing the exact distribution, in particular at large values of 
$M_{HH}$.  Similar features are observed in the $p_T$ spectrum shown in Fig.~\ref{fg:sig_pt_sm}.
Even for very small $p_T\ll m_t$, the infinite mass spectrum does not reproduce the distribution accurately, although the transverse momentum distribution is well described when including the ${\cal O} (m_t^{-4})$ terms.  However, for $p_T>m_t$, the results from the heavy mass expansion drastically fail to approximate the exact distributions.  A similar behaviour has been observed for the differential cross section $d \sigma/dp_T$ in higher order corrections to single Higgs 
production~\cite{Harlander:2012hf}.

\subsubsection{Non-Standard Model  bottom  quark Yukawa coupling}
\label{sec23_SM_nonstandard_Yb}
\begin{figure}[tb]
\begin{center}
\includegraphics[scale=0.6]{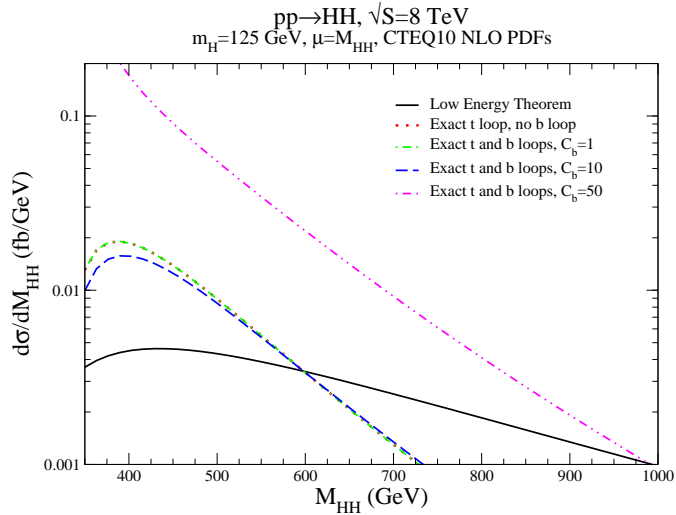}
\vskip .05in
\caption{Invariant mass distribution for Higgs pair production at $\sqrt{S}=8$~TeV, in the infinite top mass approximation
(solid black), with the full dependence on $m_t$, but no $b$ quark contribution, (red dotted) and including bottom-quark 
effects for increasing values 
of the Higgs-bottom quark Yukawa coupling (dashed lines).} 
\label{fg:mhh_b}
\end{center}
\end{figure}

We briefly discuss the role of the bottom quark loops which are omitted when using the low energy theorems.
In Fig.~\ref{fg:mhh_b}, we show the exact kinematic distribution for double Higgs production in the 
Standard Model, along with the result of the low energy theorem.
The 
bottom quark contribution is negative but negligible in the Standard Model ($C_b = {y_{bb} \over m_b } =1$ 
is a rescaling 
factor of the bottom Yukawa coupling $y_{bb}$ with respect to the Standard Model). 
The result of the destructive interference between the top and bottom quark loops remains small even when the 
bottom Yukawa is scaled up by a factor of $10$. Only enhancements by factors as large as 50 cause the 
bottom loops to dominate and give 
significant deviations. 
In the Standard Model
(with CT10 NLO PDFs and $\mu=M_{HH}$), at $\sqrt{S}=8$~TeV, the infinite mass approximation for the two Higgs cross section
is about $70\%$ of the exact two Higgs cross section.  This remains roughly true if $C_b$ is increased to $\sim 10$.
However,  if the $b$ quark Yukawa coupling is increased by a factor of $50$,
this ratio goes to $9$, and the low energy theorem is wildly inaccurate.

\subsubsection{Additional heavy quarks}
\label{sec24_ggHH_BSM}
A simple extension of the Standard Model with additional quarks of charge ${2\over 3}$ 
which can mix with the Standard Model like top occurs in many new physics scenarios, 
for example little Higgs~\cite{ArkaniHamed:2002qy,ArkaniHamed:2002pa,Low:2002ws,Perelstein:2003wd,Chang:2003un}
and composite Higgs~\cite{Giudice:2007fh,Gillioz:2012se,Espinosa:2010vn,Grober:2010yv,Hill:1991at,
Hill:2002ap,Dobrescu:1997nm,Chivukula:1998wd,He:1999vp,
Agashe:2004rs,Agashe:2006at} models. 
There can also be new heavy charge $-{1\over 3}$  quarks~\cite{Bamert:1996px,delAguila:2000rc} 
and the formulae in this 
section apply to both cases. 
We will take the new quarks to be in the fundamental representation of the color group. 
For an 
overview of the latest lower bounds on the masses of the additional quarks, 
see for example Refs.~\cite{Dawson:2012di, Okada:2012gy}. Note however that the experimental 
analyses always assume the new quarks to decay entirely either through $W$ or though $Z$. 
This is not the case in our models, and the experimental limits are therefore 
weakened~\cite{Cacciapaglia:2010vn,AguilarSaavedra:2002kr,Cacciapaglia:2011fx}.

\begin{figure}[tb]
\subfigure{
      \includegraphics[width=0.3\textwidth,clip]{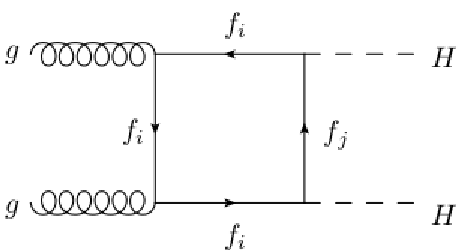}
} \quad \quad
\subfigure{
      \includegraphics[width=0.3\textwidth,clip]{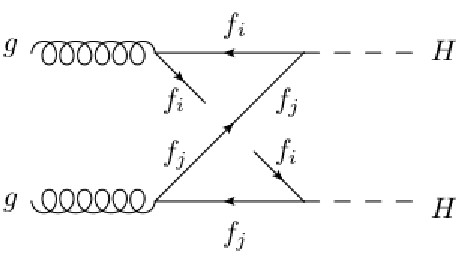}
}
\caption{Additional Feynman diagrams contributing to $gg \to HH$ in models with new heavy quarks 
coupling to the Higgs boson through non-diagonal Yukawa interactions.}
\label{fig:gghh_feynmann_bsm}
\end{figure}

In addition to the diagrams of Fig.~\ref{diags:gghh_topologies}, where any of the 
heavy quarks can be running in the loop, the double Higgs production receives contributions also from the 
mixed diagrams with two different quarks of Fig.~\ref{fig:gghh_feynmann_bsm}. 
The mass terms and the interactions  with a Higgs boson of a pair of mass eigenstate quarks $f_i, f_j$ (of the same charge)  
are
\bea
	-\lag_H &=& \sum_{i,j}  \overline{f}_{i,L} \left(
			m_i \delta_{ij} + {y_{ij} \over v} H \right) f_{j,R} + {\rm h.c.} 
\nn &=&
	 \sum_{i,j} 
	\overline{f}_i \left( m_i  \delta_{ij} + {Y_{ij} + \gamma_5 A_{ij} \over v} H \right) f_j \; ,
\label{eq:Lhm_general}
\eea
with 
\beq
\label{eq:defYA}
	Y_{ij}  = \frac{y_{ij} + y_{ji}^{*}  }{2} \qquad, \quad
	A_{ij}  = \frac{y_{ij} - y_{ji}^{*}  }{2} \;.
\eeq
We  consider real couplings. Therefore $Y_{ij} = Y_{ji} $ and  $A_{ij} = -A_{ji} $,
and only the terms involving two different quarks $f_i$ and $f_j$ contain 
pseudo-scalar couplings,
\begin{equation}
	-{\cal L}_H=\sum_i {Y_{ii}\over v} {\overline f}_{i} H f_{i} + 
		\sum_{i \neq j} {\overline f}_{i}\biggl({Y_{ij}+\gamma_5A_{ij}\over v}\biggr) H f_{j} \; .
\end{equation}
In the Standard Model $Y_{ii} = m_i$ and $A_{ij} = 0$. 

For arbitrary  masses $m_i$ and $m_j$,
\begin{eqnarray}
	F_1^{tri}(s,t,u,m_i^2,m_j^2) &=&
	{Y_{ii} \over m_i}F_1^{tri}(s,t,u,m_i^2)+{Y_{jj}\over m_j} F_1^{tri}(s,t,u,m_j^2) \nn
      &=&{4 m_H^2\over s-m_H^2}s \Tr\left(y{\cal M}^{-1}\right)+{\cal O}\biggl({s^2\over m^4}\biggr),
\label{eq:F1Triag}
\end{eqnarray}
where $y$ and ${\cal M}$ are the Yukawa and the heavy quark mass matrices from Eq.~\ref{eq:Lhm_general}. 
For the box topologies, the leading terms in the large quark mass expansion are
\begin{eqnarray}
\label{eq:F1F2general}
	F_1^{box}(s,t,u,m_i,m_j)&=&
		- {4\over 3} s \left\{
				  { Y_{ii}^2 \over m_i^2 } + { Y_{jj}^2\over m_j^2} 
				+{2 \left( Y_{ij}^2 - A_{ij}^2 \right) \over m_i m_j} 
			\right\}
		+ {\cal O}\biggl( {s^2\over m^4}\biggr) 
		\nn	
		&=&
		- {4\over 3} s\, \Tr\left(y{\cal M}^{-1}y{\cal M}^{-1}\right)+{\cal O}\biggl( {s^2\over m^4}\biggr) \; ,
		\nn
		F_2(s,t,u,m_i,m_j)&=& 
		{\cal O}\biggl( {s^2\over m^4}\biggr) \;.
		\label{eq:F1boxLET}
\end{eqnarray}
The relative minus sign between the vector and axial contributions comes from Eq.~\ref{eq:defYA}.  

Although the leading terms of the triangle and box diagrams were calculated in the diagonal mass basis, 
the cyclicity of the trace and the fact that both ${\cal M}$ and $y$ rotate according to the same 
unitary transformations allow one to cast the results in Eqs.~\ref{eq:F1Triag} and~\ref{eq:F1boxLET} 
into a basis independent form. Hence the Yukawa and mass matrices can be evaluated both in the 
mass basis, where ${\cal M}$ is diagonal, and in the current basis. In the current basis, $y={\partial {\cal M} \over \partial v}$. 
The infinite mass limit of both the triangle and box diagrams can also be obtained 
via the low energy theorems~\cite{Kniehl:1995tn,Low:2009di}.

In our calculations in Sections~\ref{sec41_top_partner} and~\ref{sec42_mirror_fermions},
 we retain the full dependence of the leading order amplitude on the quark masses. 
However, for small mass splitting $\delta\equiv m_j^2 - m_i^2$ the sub-leading terms have a simple 
and useful form,  
\begin{eqnarray}
	F_1^{box}(s,t,u,m_i^2,\delta)&=&
		{Y_{ii}^2+Y_{jj}^2+2Y_{ij}^2\over m_i^2}		F_1^{box}(s,t,u,m_i^2)
		+ {4 \over 3} s { Y_{jj}^2+Y_{ij}^2 \over m_i^2}
		{\delta \over m_i^2} \left[1+{7\over 10}{m_H^2\over m_i^2} \right]
\nn & & 
		+{8\over 3}s {A_{ij}^2\over m_i^2}
		\left[1 + {15 m_H^2 - 4 s   \over 60 m_i^2} - {\delta \over 2 m_i^2}
		\right]
		+ {\cal O} \biggl({s^2\over  m_i^4},{\delta^2\over  m_i^4} \biggr)  \; ,
\nonumber \\
	F_2(s,t,u,m_i^2,\delta)&=&
		{Y_{ii}^2+Y_{jj}^2+2Y_{ij}^2\over m_i^2}
		F_2(s,t,u,m_i^2)
	+ s {Y_{jj}^2+Y_{ij}^2 \over m_i^2}
		{\delta \over m_i^2}\left({22\over 45}{p_T^2\over m_i^2}\right)
	\nonumber \\
	&&
	-{2\over 3} s  {A_{ij}^2 \over m_i^2} {p_T^2\over m_i^2}
	+ {\cal O} \biggl({s^2\over  m_i^4}, {\delta^2\over  m_i^4} \biggr)  \, .
\end{eqnarray}

Following~\cite{Falkowski:2007hz}, we consider the infinite quark mass limit of these results and recast them 
into a convenient form for the calculation of the amplitudes for single and double Higgs production in models with extended quark
sectors with respect to the Standard Model amplitudes. 
In the infinite mass approximation, the leading order amplitudes can be written as (Eqs.~\ref{eq:F1Triag},~\ref{eq:F1boxLET}) 
\beq
	{A}_{gg \to H}  \propto \Tr\left(y {\cal M}^{-1} \right)\qquad, \quad
	{A}_{gg \to HH}^{box}  \propto  \Tr\left(y {\cal M}^{-1} y {\cal M}^{-1} \right) \; ,
\eeq
where the  omitted proportionality terms do not depend on the masses and Higgs couplings of the quarks. 
In the Standard Model, 
$y_{tt} = m_t$. The amplitudes only depend on the omitted proportionality factors, which therefore cancel when taking the 
ratio to the Standard Model result:
\bea
\label{eq:Rs_for_H_prodn}
	R_{gg \to H} &\equiv & {{A}_{gg \to H} \over A_{gg \to H}^{SM} } = \Tr\left(y {\cal M}^{-1} \right) 
	= {\partial \over \partial v} \left( \log \det {\cal M} \right) \; , \label{falk}\\
	R_{gg \to HH}^{box} & \equiv & {{A}_{gg \to HH}^{box} \over A_{gg \to HH}^{box, SM} } 
		= \Tr\left(y {\cal M}^{-1} y {\cal M}^{-1} \right) \label{2low}
\eea
In Eq.~\ref{falk} we used the relation $y={\partial {\cal M} \over \partial v}$~\cite{Falkowski:2007hz}. 
Eq.~\ref{2low}  is equivalent to the result of Ref.~\cite{Gillioz:2012se}.

\section{Examples}
\label{sec4_Examples}

\subsection{Singlet top partner}
\label{sec41_top_partner}

We are interested in examining possible large effects in two Higgs production from gluon fusion
in models which are consistent with precision electroweak measurements and the observed rate
for single Higgs production.
Topcolor models~\cite{Hill:1991at,Hill:2002ap}, top condensate 
models~\cite{Dobrescu:1997nm,Chivukula:1998wd,He:1999vp,He:2001fz}, 
and little Higgs models~\cite{ArkaniHamed:2002qy,Low:2002ws,Perelstein:2003wd,
Chang:2003un,Chen:2003fm,Hubisz:2005tx,Han:2005ru} all
contain a charge ${2\over 3} $ partner of the top quark. We consider a general case
with a vector $SU(2)_L$ singlet fermion, ${\cal T}^2$, which is allowed
to mix with the Standard Model like top quark, 
${\cal T}^1$~\cite{ Lavoura:1992qd,Maekawa:1995ha,Popovic:2000dx,
AguilarSaavedra:2002kr,Cacciapaglia:2010vn,Dawson:2012di}. 
The fermions are,
\beq
	\psi_L=\left( \begin{matrix} {\cal T}^1_L\\{\cal B}^1_R\end{matrix}\right) \, ,
	\quad {\cal T}^1_R, {\cal B}^1_R \; ; {\cal T}^2_L, {\cal T}^2_R
\; .
\eeq
Following the notation of~\cite{Dawson:2012di}, 
the mass eigenstates  are $t,T$ and $b={\cal{ B}}^1$ 
(where $t,b$ are the observed top and bottom quarks), and can be found by
the rotations 
\begin{equation}
\chi_{L,R}^t\equiv
\left( \begin{matrix}
t_{L,R}\\
T_{L,R}\end{matrix}\right)
\equiv  U_{L,R}^t
\left(\begin{matrix}
{\cal{T}}^1_{L,R}\\{\cal{T}}^2_{L,R}
\end{matrix}
\right)\; .
\end{equation}
The chirality projectors are $P_{L,R}\equiv{\
1\mp\gamma_5\over 2}$ and the mixing matrices $U_L^t, U_R^t$ are unitary 
and parameterized as,
\begin{eqnarray}
U_L^t&=&
\left(\begin{matrix}
\cos\theta_L& -\sin\theta_L\\
\sin\theta_L& \cos\theta_L\end{matrix}
\right),\quad
U_R^t=
\left(\begin{matrix}
\cos\theta_R & -\sin\theta_R\\
\sin\theta_R & \cos\theta_R\end{matrix}
\right) \, .
\end{eqnarray}
We will abbreviate $s_L = \sin \theta_L$, $c_L = \cos \theta_L$. 

The fermion mass terms are 
\begin{eqnarray}
	-{\cal L}_{M,1}&=&
	\lambda_1 \overline{\psi}_L H {\cal B}^1_R
	+\lambda_2 \overline{\psi}_L \tilde{H}{\cal T}_R^1+
	\lambda_3 \overline{\psi}_L \tilde{H} {\cal T}^2_R+
	\lambda_4 \overline{\cal T}^2_L {\cal T}^1_R+
	\lambda_5 \overline{\cal T}^2_L {\cal T}^2_R+ \rm{h.c.}
	\nn &=&
	\overline{\chi}_L^t \biggl[U_L^t M^t_{(1)}U_R^{t \dagger}\biggr]\chi_R^t \
	+
	\lambda_1{v\over\sqrt{2}} \overline{\cal B}^1_L{\cal B}^1_R+ {\rm h.c.} \; ,
\end{eqnarray}
where
\begin{equation}
M^t_{(1)}=\left(
\begin{matrix}\lambda_2{v\over\sqrt{2}}&\lambda_3{v\over\sqrt{2}}\\
\lambda_4&\lambda_5\end{matrix}
\right) \; .
\end{equation}
Without loss of generality, the  
${\overline {\cal T}}^2_L {\cal T}^1_R$ term can
be rotated away through a redefinition of the right handed fields. The model therefore 
contains three independent parameters in the top sector, which we take to be
 $m_t, M_T$ and $\theta_L$. 
The consistency of the model with electroweak precision measurements and its 
decoupling properties have been studied in many 
works~\cite{ Lavoura:1992qd,Maekawa:1995ha,Popovic:2000dx,AguilarSaavedra:2002kr,Dawson:2012di,Okada:2012gy}. 
We will not repeat this analysis here, but use the results of Ref.~\cite{Dawson:2012di}. 
It is interesting to note that in the limit $\theta_L\sim 0$ (required by precision electroweak data),
the mass terms for the top like quark and its partner become 
\begin{eqnarray}
	\lambda_2&\simeq & {\sqrt{2}m_t\over v} \left[ 1+{s_L^2\over 2}(r-1) \right ]
	  \; ,\nn
	  \lambda_5& \simeq &M_T \left[1+{s_L^2\over 2}{1-r \over r } \right] \; ,
\end{eqnarray}
where $r = {M_T^2 \over m_t^2}$. Decoupling of the heavy quark therefore requires 
$s_L^2 \sim r^{-1}$, as it was shown in~\cite{Dawson:2012di}.

\begin{figure}
\begin{center}
\includegraphics[scale=0.6]{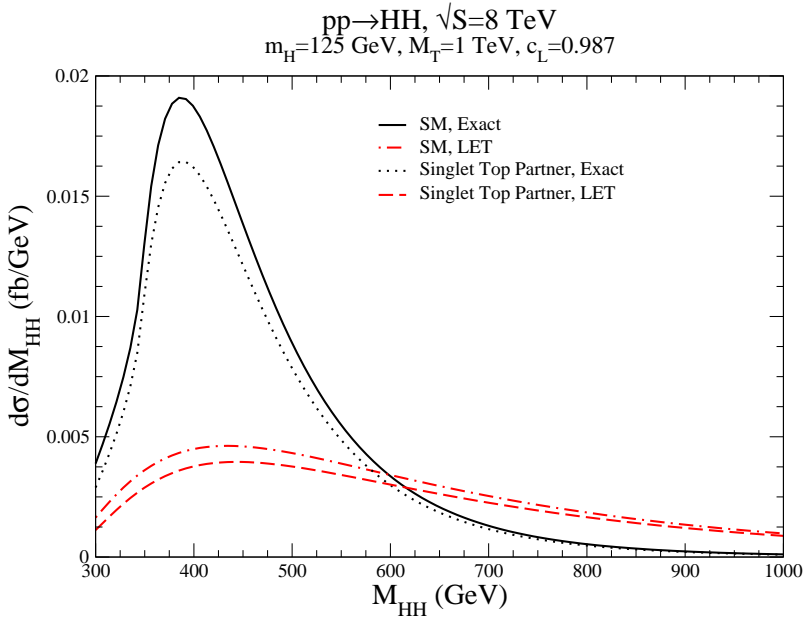}
\vskip .05in
\caption{Invariant mass distribution in the Standard Model and in the top-singlet partner model (with 
$M_T =$~1 TeV) at the $\sqrt{S} = 8$ TeV LHC.}
\label{fg:hh_singlet}
\end{center}
\end{figure}

Since we are interested in Higgs production from the quark loops, we need 
the couplings to the Higgs boson, 
\begin{equation}
-{\cal L}_{H,1}={m_t\over v}c_{tt}{\overline t}_Lt_R H
+{M_t\over v}c_{TT}{\overline T}_LT_R H
+{M_T\over v} c_{tT}{\overline t}_L T_R H
+{m_t\over v}c_{Tt}{\overline T}_Lt_R H
+{\rm h.c.} \; ,
\end{equation}
where
\beq
	c_{tt}= c_L^2 \qquad,\quad 	c_{TT}= s_L^2 \qquad,\quad 	
	c_{tT}= c_{Tt} = s_L c_L \;.
\label{singleh}
\eeq
Using Eq.~\ref{singleh} and the low energy theorems of Eqs.~\ref{eq:Rs_for_H_prodn} and~\ref{2low}, it is straightforward to 
see that the single and double Higgs production rates are the same as the Standard Model up to
corrections of ${\cal O}\left({s\over m_t^2},{s\over M_T^2}\right)$. 
These corrections are further suppressed 
by the small mixing angles allowed by the bounds from electroweak precision data~\cite{Dawson:2012di}.
Both total and differential distributions are very close to the Standard 
Model~(Fig.~\ref{fg:hh_singlet}), 
and one cannot use double Higgs production to 
obtain information about additional vector singlet quarks. Fig.~\ref{fg:hh_singlet} uses the largest mixing
angle allowed by precision electroweak data, and the reduction in the total cross section for the singlet top
partner model from the exact Standard Model result is roughly $15\%$. 
This is of similar size to the reduction in the $gg\rightarrow H$ rate found in Ref.~\cite{Dawson:2012di}.  
This model is an example of a case which will be extremely difficult to differentiate from the Standard Model.

\subsection{Mirror fermions}
\label{sec42_mirror_fermions}
As a second example, we consider a model which has a generation of heavy mirror  
fermions~\cite{He:2001tp,Erler:2010sk,Eberhardt:2010bm,Martin:2009bg,Lavoura:1992qd}.
There are four new quarks ${\cal T}^1$, ${\cal T}^2$ and ${\cal B}^1$, 
${\cal B}^2$, with charge ${2\over 3}$ and $-{1\over 3}$, respectively. The quarks are in the $SU(2)_L$ 
representations,
\begin{eqnarray}
	\psi_L^1  &=&
	\left(\begin{matrix}
		{\cal T}_L^1\\
		{\cal B}_L^1
	\end{matrix}\right)
	\, ,\;\; 
	{\cal T}_R^1\, , {\cal B}_R^1 \; ; \quad\quad 
	\psi_R^2  =
	\left(\begin{matrix}
		{\cal T}_R^2\\
		{\cal B}_R^2
	\end{matrix}\right)
	\, ,\;\; 
	{\cal T}_L^2\, , {\cal B}_L^2 \, .
\end{eqnarray}
The first set of heavy quarks has the quantum numbers of the Standard Model quarks, while ${\cal T}^2$ and ${\cal B}^2$ have
have the left- and right- handed fermion assignments reversed from those of the Standard Model. 
For simplicity, we assume there is no mixing between the heavy mirror fermions and the Standard Model 
fermions.  This assumption eliminates the need to consider limits from $Z\rightarrow b {\overline b}$~\cite{Bamert:1996px} 
and relaxes the restrictions from precision electroweak data discussed in 
Sec.~\ref{sec42b_bounds_from_ewd}\footnote{We will not explore 
UV completions of this model that can mediate the decay of the 
mirror fermions through higher-dimensional operators and prevent the new quarks from becoming stable.}.

The most general Lagrangian for the interactions of the mirror fermions with the Higgs 
doublet is,
\begin{eqnarray}
\label{eq:lag_mirror}
	-{\cal L}
	&=& 
	  \lambda_A{\overline \psi}_L^1\Phi {\cal B}_R^1 +\lambda_B{\overline \psi}_L^1 {\tilde \Phi} {\cal T}_R^1
	  +\lambda_C{\overline \psi}_R^2\Phi {\cal B}_L^2 +\lambda_D{\overline \psi}_R^2 {\tilde \Phi} {\cal T}_L^2
	\nn &&
	+\lambda_E{\overline \psi}_L^1 \psi_R^2 + \lambda_F{\overline {\cal T}}_R^1 {\cal T}_L^2 
	+\lambda_G {\overline {\cal B}}_R^1 {\cal B}_L^2 + {\rm h.c.}
\nn &=&
	{\overline {\chi}}_L^t \biggl[
		U_L^{t}  {\cal M}_{U} U_R^{t \dagger}  \biggr]	\chi_R^t
	+ {\overline {\chi}}_L^b \biggl[
		U_L^{b}	{\cal M}_{D}U_R^{b \dagger}\biggr]\chi_R^b 
		+{\rm h.c.} \,.
\end{eqnarray}
The mass eigenstates $\chi_P^q$ $(P=L,R\, ;\, q=t,b)$ are obtained through unitary rotations 
\bea
	U_P^q
	&=&
	\begin{pmatrix}
			\cos\theta_P^q	&	-\sin\theta_P^q	\\
			\sin\theta_P^q	&	\cos\theta_P^q
	\end{pmatrix} \, ,
\eea 
and the mass matrices are
\begin{eqnarray}
	{\cal M}_U
	&=&
	\begin{pmatrix}
		\lambda_B {v\over \sqrt{2}} & \lambda_E	\\
		\lambda_F & \lambda_D{v\over \sqrt{2}}
	\end{pmatrix}
	\, , \qquad
	{\cal M}_D =  
	\begin{pmatrix}
		\lambda_A {v\over \sqrt{2}} & \lambda_E	\\
		\lambda_G & \lambda_C{v\over \sqrt{2}}
	\end{pmatrix} \,.
\end{eqnarray}
We will denote the two top- like and the two bottom- like mass eigenstates as $T_1, T_2$ 
and $B_1, B_2$ respectively. 
The Lagrangian parameters $\lambda_i$ can be 
expressed in terms of the physical quark masses and the mixing angles. We report these 
relations in Appendix~\ref{app1_ew_mirror}. 

Since all the quarks have different quantum numbers, it is not possible 
to rotate away any parameter in the Lagrangian. However, the $SU(2)$ symmetry 
requires that
\beq
	{\cal M}_{U,12} = {\cal M}_{D,12} \, ,
\eeq
and therefore
\beq
\label{eq:reln_masses_angles_LR}
	   M_{T_2} \cos \theta^t_R \sin \theta^t_L 
	- M_{T_1} \cos \theta^t_L \sin \theta^t_R 
	= 
	   M_{B_2} \cos \theta^b_R \sin \theta^b_L 
	- M_{B_1} \cos \theta^b_L \sin \theta^b_R \, .
\eeq
This relation can be written as 
\beq
\label{eq:reln_masses_angles_PM}
	\biggl[M_{T_2}+M_{T_1}\biggr]\sin\theta^t_-
	+\biggl[M_{T_2}-M_{T_1}\biggr]\sin\theta^t_+
	=
	\biggl[M_{B_2}+M_{B_1}\biggr]\sin\theta^b_-
	+\biggl[M_{B_2}-M_{B_1}\biggr]\sin\theta^b_+ \; ,
\eeq
where $\theta^{t(b)}_\pm = \theta^{t(b)}_L\pm\theta^{t(b)}_R$.

The couplings of the fermion mass eigenstates to the Higgs boson are 
\bea
	-\lag ^H_M 
	& = &
	{c_{T_1 T_1}\over 2v} {\overline T}_{1L} T_{1R} \,H
	+ { c_{T_2 T_2}\over 2 v} {\overline T}_{2L} T_{2R}\,H 
	+ {c_{T_1 T_2} \over 2v} {\overline T}_{1L} T_{2R}\,H
	+ {c_{T_2 T_1} \over 2v} {\overline T}_{2L} T_{1R}\,H + \nn
&&
	{c_{B_1 B_1} \over 2v} {\overline B}_{1L} B_{1R} \, H
	+ {c_{B_2 B_2} \over 2v}{\overline B}_{2L} B_{2R}\,  H
	+ {c_{B_1 B_2} \over 2v} {\overline B}_{1L} B_{2R} \, H
	+ {c_{B_2 B_1} \over 2v} {\overline B}_{2L} B_{1R} \,H + {\rm h.c.}\, ,\nn
\eea%
where
\bea
\label{eq:Ycoupl_mirror}
		c_{T_1 T_1} & = & 
			M_{T_1} \left[1 + \cos \left( 2 \theta_L^t \right)\cos \left( 2 \theta_R^t \right) \right] + 
			M_{T_2} \sin \left( 2 \theta_L^t \right) \sin \left( 2 \theta_R^t \right) \nn
			&=&
			2M_{T_1}\left[\cos^2\theta^t_- -\frac{M_{T_1}-M_{T_2}}{2M_{T_1}}\left(\sin^2\theta^t_+-\sin^2\theta^t_-\right)\right] \, ,
\nn
		c_{T_1 T_2} & = & 
			M_{T_1} \cos \left( 2 \theta_L^t \right)\sin \left( 2 \theta_R^t \right)
			-M_{T_2} \cos \left( 2 \theta_R^t \right)  \sin \left( 2 \theta_L^t \right)  \nn
		& = &
		\frac{M_{T_1}-M_{T_2}}{2}\sin(2\theta^t_+)-\frac{M_{T_2}+M_{T_1}}{2}\sin(2\theta^t_-)\, ,
	\nn
	( c_{T_2 T_2} , c_{T_2 T_1}  ) &=& (c_{T_1 T_1} ,c_{T_1 T_2} ) {\rm \; with \;}
	M_{T_1} \leftrightarrow M_{T_2} , \theta^t_\pm \rightarrow - \theta^t_\pm \;.
\eea
Similar expressions hold in the bottom sector. 

The couplings to the electroweak gauge bosons that are needed for the computation 
of the Peskin--Takeuchi parameters (Sec.~\ref{sec42b_bounds_from_ewd}) are reported in the Appendix.

\label{sec42a_LET_mirror}
\subsubsection{Higgs production using  low energy theorems in the mirror model}

For single Higgs production through top quark and mirror fermion loops, 
the low energy theorem of Eq.~\ref{falk} yields
\begin{eqnarray}
	A_{gg\to H}&=&
	A_{gg\to H}^{SM}
	\left(1+{c_{T_1 T_1 }\over 2 M_{T_1}}
	+{c_{T_2 T_2}\over 2 M_{T_2}}
	+{c_{B_1 B_1 }\over 2 M_{B_1}}
	+{c_{B_2 B_2}\over 2 M_{B_2}}\right) 
	\equiv
	A_{gg\to H}^{SM}
	\left(1+\Delta\right)\; ,
\end{eqnarray}
where we introduce the fractional difference $\Delta$ of the single Higgs amplitude from that of the Standard Model.  

Both for simplicity and because one expects large corrections to the oblique parameters 
for a large mass splitting within each chiral 
doublet, we assume  
$M_{T_1}= M_{B_1}=M$ and $M_{T_2}=M_{B_2}=M(1+\delta)$. 
In this limit,
\beq
	\label{SingHiggs.EQ}
	A_{gg\rightarrow H}	=	A_{gg\rightarrow H}^{SM}
	\left\{1+4-\frac{1}{1+\delta}\left[(2+\delta)\sin\theta_-^t-\delta\sin\theta_+^b\right]
		\left[(2+\delta)\sin\theta^b_-+\delta\sin\theta^b_+\right]
	\right\} \, ,
\eeq
where we impose (see Eq.~\ref{eq:reln_masses_angles_PM}) 
\beq
	\label{eq:special_reln_masses_angles_PM}
	\left( 2 + \delta \right) \sin\theta^t_- +\delta \sin\theta^t_+
	=
	\left( 2 + \delta \right)\sin\theta^b_- +\delta \sin\theta^b_+ \; .
\eeq

Given the recent observations at the LHC, we are interested in the case when 
\mbox{$A_{gg \to H}\sim A_{gg \to H}^{SM}$}. One simple way to recover this
limit is to have 
\beq
	\theta_-^t \sim  {\pi\over 2} \, , \quad \theta_-^b  \sim  {\pi\over 2}\, ,
\label{ang_SM}
\eeq
which for single production gives\footnote{This relation holds for small $\delta$. For $\delta = 0$, 
Eq.~\ref{eq:reln_masses_angles_PM} requires $\sin \theta^t_- = \sin \theta^b_-$, and 
\mbox{$A_{gg\to H}=A_{gg\to H}^{SM}\left(1+4 \cos^2 \theta^b_- \right)$}. 
This result can be easily understood from the Yukawa couplings, 
$ c_{T_1 T_1} = c_{T_2 T_2 } = M \cos^2 \theta^b_- $ and $c_{T_2 T_1} = -c_{T_1 T_2} = {M \over 2} \sin(2 \theta^b_-) .
$   
Also in this case, the $gg\rightarrow H$ rate is identical to  the Standard Model rate  for $\theta^b_- = {\pi \over 2}$. 
}
\begin{eqnarray}
	A_{gg \to H}		&\sim &
	A_{gg\to H}^{SM}
	\left\{1-\frac{\delta^2}{1+\delta}\cos^2\theta^b_+ \right\}\,=A_{gg \to H}^{SM}
	\left\{1-\frac{\delta^2}{1+\delta}\sin^2(2\theta^b_R) \right\} \; .
\label{SingHiggsLim.EQ}
\end{eqnarray}
To get the Standard Model result for $gg \to H$ further requires either $\delta\sim 0$ 
or $\theta_R^b \sim  \theta_R^t \sim 0$, 
where the constraint on the right-handed mixing angle in the top sector arises from 
Eq.~\ref{eq:special_reln_masses_angles_PM}. 

The result of Eq.~\ref{SingHiggsLim.EQ} can be understood by inspecting the Yukawa couplings in the 
limit  $\theta_-^{t,b} = {\pi\over 2}$:
%
\begin{eqnarray}
	c_{T_1 T_1} = -c_{T_2 T_2} =& -M \delta \cos^2(\theta^t_+) & = -M \delta \sin^2(2 \theta^t_R) \; ,\nn
	c_{T_1 T_2}= \phantom{-}c_{T_2 T_1}  = & - \frac{M \delta}{2} \sin(2 \theta^t_+) & = 
		\phantom{-} \frac{M \delta}{2} \sin(4 \theta^t_R) \; .
\end{eqnarray}
Similar relations hold for the charge $-{1\over 3}$ sector.  
Hence, for $\delta\sim0$ or $\theta^{t,b}_R\sim0$ the diagonal Yukawa 
couplings go to zero and only the top quark, with its Standard Model Yukawa coupling, contributes to single Higgs production. 
The off-diagonal couplings of the mirror fermions to the Higgs boson are slightly less suppressed, and could induce 
deviations in the double Higgs rate from that of the Standard Model.

From the low energy theorem of Eq.~\ref{2low}, the box contributions to $gg\rightarrow HH$  production (including top quark loops)
can be estimated,
\bea
\label{eq:F1boxmirror}
F_1^{box}&\equiv &
F_1^{box, SM} \left(1+\Delta_{box} \right) \; ; \nn
\Delta_{box} &=&
{c_{T_1 T_1}^2\over 4M_{T_1}^2}
+{c_{T_2T_2}^2\over 4M_{T_2}^2}
+{c_{B_1B_1}^2\over 4M_{B_1}^2}
+{c_{B_2B_2}^2\over 4M_{B_2}^2}
+{c_{T_1 T_2 }c_{T_2 T_1} \over 2M_{T_1}M_{T_2}}+{c_{B_1 B_2}c_{B_2 B_1}\over 2 M_{B_1}M_{B_2}} \nn
&=&
4 + \frac{3}{2} \frac{ \alpha_1^2 - \alpha_2^2 +\alpha_3^2 - \alpha_4^2 }{1+\delta}
+ \frac{1}{4} \frac{ (\alpha_1^2 - \alpha_2^2)^2 + (\alpha_3^2 - \alpha_4^2)^2 }{(\delta+1)^2} 
\; ,
\end{eqnarray}
where we defined 
\bea
	\alpha_1 &=&  \delta \sin \theta^b_+ + (2+\delta) ( \sin \theta^b_- - \sin \theta^t_-) \; ,
	\nn
	\alpha_2 &=&  (2+\delta) \sin \theta^t_- \quad , \qquad
	\alpha_3 =  \delta \sin \theta^b_+ \quad , \qquad
	\alpha_4 =  (2+\delta) \sin \theta^b_- \; .
\eea	
For $\theta^{t,b}_- \sim {\pi \over 2}$, Eq.~\ref{eq:F1boxmirror} yields\footnote{In the exact $\delta=0$ limit the result reads
$
	F_1^{box}=
	F_1^{box, SM} \left[1-4 \cos^2 \theta^b_- + 8 \cos^4 \theta^b_- \right].$
}
\beq
	\Delta_{box} =
	\ {\delta^2 \over 1 + \delta } \cos^2 \theta^b_+ + {\delta^4 \over 2 (1 + \delta)^2 } \cos^4 \theta^b_+ 
		\; .
\eeq
Note that $F_2^{box}$ does not contribute in the infinite fermion mass limit.
The terms proportional to  $\cos^2 (\theta^b_+) $ come from the contributions of the off-diagonal fermion-Higgs
 couplings. 
For this simple choice of parameters, the same term governs the deviations from the Standard Model 
both in single and double Higgs production. 

\begin{figure}[tb]
	\begin{center}
	\includegraphics[scale=0.6]{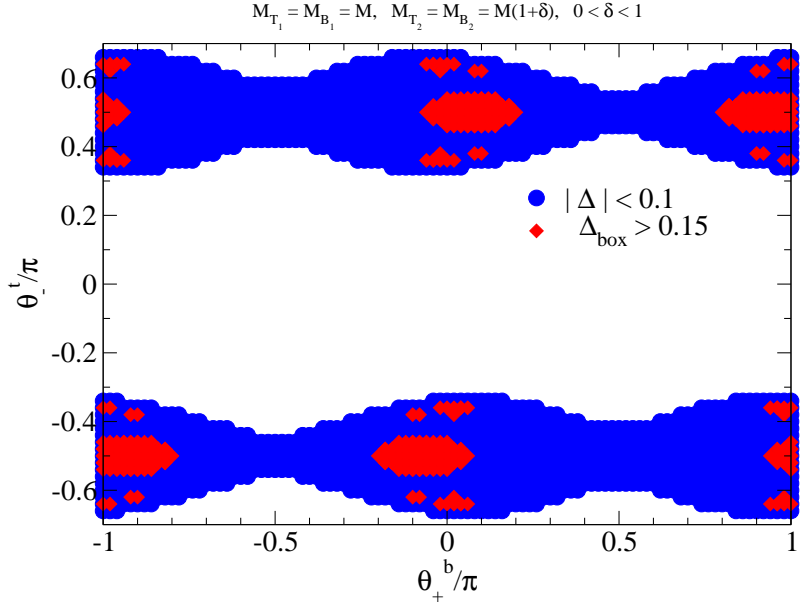}	\vskip .05in
	\caption{Allowed regions in the $\theta^t_-, \theta^b_+$ parameter space where deviations, $\Delta$,
	 from the Standard Model $gg\rightarrow H$ amplitude  are  below $10\%$ and the  mirror fermion masses
	 satisfy $0<\delta<1$. 
	The other two angles are fixed through Eqs.~\ref{eq:special_reln_masses_angles_PM} and~\ref{eq:bm_for_Delta}. 
	The red diamonds denote regions where the $gg\rightarrow HH$ amplitude from the box topology deviates from the Standard 
	Model by more than $15\%$.  
	}
	\label{fg:Scan_thTm_thBp}
	\end{center}
\end{figure}

We are interested in determining how large a deviation from the Standard Model \mbox{$gg\rightarrow HH$} rate is possible 
with a minimal deviation in the $gg\rightarrow  H$ rate.  With the assumption of no mass splitting within the mirror doublets, 
there are five independent parameters: the mass scale $M$, which drops out in the heavy mass limit for the Higgs 
production rates, the mass 
splitting between families, $\delta$, and three angles. Using Eq.~\ref{SingHiggs.EQ}, we replace 
one of the angles with the fractional deviation $\Delta$ of the $gg\rightarrow H$ amplitude  from that of  the Standard Model, 
\begin{equation}
\label{eq:bm_for_Delta}
\sin\theta^b_- = \frac{1}{2+\delta}\biggl\{\frac{(4-\Delta)(1+\delta)
}{(2+\delta)\sin\theta^t_--\delta\sin\theta^b_+}-\delta\sin \theta^b_+\biggr\} \; .
\end{equation}
We require this deviation to be within $10\%$ and the mass splitting $\delta$ 
between the two mirror families not to be too large ($0<\delta<1$), since we expect electroweak observables 
to put severe bounds on $\delta$.  Under these constraints, we perform a scan over $\delta, \Delta, \theta^t_-$ and 
$\theta^b_+$. The values of these parameters for which Eqs.~\ref{eq:special_reln_masses_angles_PM} and~\ref{eq:bm_for_Delta} 
yield real solutions for $\theta^t_+, \theta^b_-$ are represented by the blue dots in Fig.~\ref{fg:Scan_thTm_thBp}. 
The red diamonds represent regions where the 
difference %
$\Delta_{\rm box}$ 
in the double Higgs amplitude from the box topology is larger than $15\%$. 

In the following, we fix $\theta^t_- = {\pi \over 2}$ in order to focus on a region with large $\Delta_{\rm box}$, and 
analyse how double Higgs production depends on $\theta^b_+$ and $\delta$ for a Standard Model $gg\rightarrow
H$ amplitude,  $\Delta = 0$,  and for $\pm 10\%$ deviations from it, $\Delta = \pm 0.1$. 
\begin{figure}[tb]
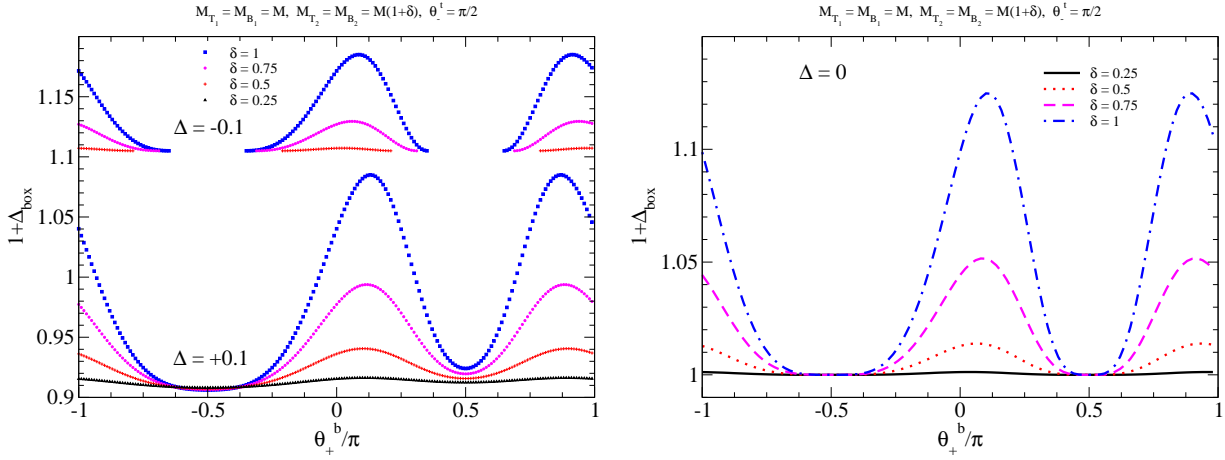

\subfigure{
	\includegraphics[scale=0.45]{TwoHiggsAmp_Delt_pm01.eps}
}
\subfigure{
	\includegraphics[scale=0.45]{TwoHiggsAmp.eps}
}
\caption{Deviations from the Standard Model box amplitude, $F_1^{box}$,  as a function of 
	$\theta^b_+$ for $\theta^t_- = {\pi \over 2}$, $M = 800$~GeV
	and four different values of the fractional mass 
	difference $\delta$ of the two mirror families, for a $10\%$ deviation in the single 
	Higgs production amplitude (left plot) and for the same $gg\rightarrow H$ amplitude
	 as in the Standard Model (right plot). 
	The blank regions on the curves are not allowed for $\theta^t_+, \theta^b_-$ to be real.
}
\label{fg:TwoHiggsAmp_Delt_pm01_0}
\end{figure}
This analysis is shown in Fig.~\ref{fg:TwoHiggsAmp_Delt_pm01_0} for a heavy mass scale 
$M = 800$~GeV. To qualitatively understand the features of these plots, one can consider the 
limit of small deviations from the Standard Model single Higgs amplitude and small 
family splitting $\delta$, 
\beq
	\Delta_{box}=
	\delta^4 \cos^4 \theta^b_+ \left[ {1 \over 2}  - \delta ( 1 - \sin \theta^b_+) \right]
		+ \Delta \left[ -1 + \delta^2  \cos^2 \theta^b_+ 
	+ {\cal O}(\delta^3)\right]
	+{\cal O} (\Delta^2, \delta^6) \;.
	\label{eq:deltabox_smalldD}
\eeq
For almost degenerate mirror fermions ($\delta\sim 0$) and small deviations in single Higgs production from
the Standard Model case, (which occurs when $\theta^b_+ = \pm
{\pi \over 2}$), the dominant term is $\Delta_{box}\sim -\Delta$. When single Higgs production is suppressed, 
double Higgs
production is always enhanced, while for a slightly enhanced Higgs single production rate, 
double production can also be suppressed. For $\Delta = 0$ and small $\delta$, double Higgs 
production is also enhanced. In all cases, the minimal deviations from Standard Model double Higgs production 
occurs exactly at $\theta^b_+ = \pm{\pi \over 2}$, while the maximum deviation is at 
\beq
	\theta^b_+  = \arccos \left( 2 {  \sqrt{1+\delta} \over 2+ \delta}  \right)  
	+ {\sqrt{1+\delta} \over 2 \delta} \Delta + {\cal O}(\Delta^2)  \; .
\eeq

Finally, we note that the results of this section can be written in terms of an effective Lagrangian, which for $\delta=0$ is 
\begin{equation}
{\cal L}_{eff}={\alpha_s\over 12 \pi } G_{\mu\nu}^aG^{a,\mu\nu} 
\left[(1+ 4\cos^2\theta_-^b) \frac{H}{v}
-(1-4\cos^2\theta_-^b+8\cos^4\theta_-^b){H^2\over 2 v^2}\right]
\; .
\end{equation}

\label{sec42b_bounds_from_ewd}
\subsubsection{Bounds from electroweak precision data}
The new mirror quarks carry electroweak charges, and therefore contribute to the self -
energies of the electroweak gauge bosons~\cite{He:2001tp,Maekawa:1995ha,Lavoura:1992np}. 
A convenient way to parametrize these 
effects is through the Peskin--Takeuchi parameters~\cite{Altarelli:1990zd, Peskin:1991sw}, 
\begin{eqnarray}
	\alpha\Delta S_F &=&
	{4 s_W^2 c_W^2\over M_Z^2}
	\biggl\{ 
	\Pi_{ZZ}(M_Z^2)- \Pi_{ZZ}(0)-\Pi_{\gamma\gamma}(M_Z^2)
-{c_W^2-s_W^2\over c_W s_W}
\Pi_{\gamma Z}(M_Z^2)
\biggr\} \; ,\nn
%
\alpha \Delta T_F &=&
{ \Pi_{WW}(0)\over M_W^2}
-{\Pi_{ZZ}(0)\over M_Z^2} \; ,\nn
\alpha \Delta U_F&=& 4 s_W^2\biggl\{
{ \Pi_{WW}(M_W^2)-\Pi_{WW}(0)\over M_W^2} 
-c_W^2\biggl({ \Pi_{ZZ}(M_Z^2)-\Pi_{ZZ}(0)\over M_Z^2}\biggr)
\nonumber \\
&&-2 s_W c_W
{ \Pi_{\gamma Z}(M_Z^2)
\over M_Z^2}
-s_W^2 { \Pi_{\gamma \gamma}(M_Z^2)\over M_Z^2}\biggr\} \, ,
\label{sdef}
\end{eqnarray}
where $\Pi_{XY}(p^2)$ denotes the transverse part of the vacuum polarization amplitude evaluated at 
momentum $p^2$ and $c_W^2={M_W^2\over M_Z^2}=1-s_W^2$. The couplings of the mirror fermions 
to the electroweak gauge bosons are reported in the Appendix. 

We use the fit to the electroweak precision data given in Ref.~\cite{Baak:2012kk}, 
\beq
	S = 0.03 \pm 0.10 \quad, \qquad 
	T = 0.05 \pm 0.12 \quad, \qquad  
	U = 0.03 \pm 0.10 \;,
	\label{delts}
\eeq
with correlation coefficients 
\begin{eqnarray}
\rho_{ij}=\left(
\begin{array}{lll}
1.0 & 0.89 & -0.54\nonumber \\
0.89 & 1.0 & -0.83 \nonumber\\
-0.549 & -0.83 & 1.0
 \end{array}
\right)\, .
\end{eqnarray}
The reference Higgs and top-quark masses are
$m_{H,{\rm ref}}=126$~GeV and $m_{t,{\rm ref}} = 173$~GeV. 
We use $m_H=125$~GeV and so we need to account also 
for the Higgs contributions to the electroweak parameters. Up to terms of 
${\cal O} (M_Z^2/m_H^2)$, they read 
\beq
	\Delta S_H= {1\over 12 \pi}
\log\biggl(	{m_H^2\over m_{H,{\rm ref}}^2}	\biggr) \quad ,\quad
	\Delta T_H= -{3\over 16 \pi c_W^2}
\log\biggl( {m_H^2\over m_{H,{\rm ref}}^2}\biggr) \quad,\quad
	\Delta U_H= 0 \;.
\eeq
The $\Delta \chi^2$ is defined as
\begin{equation}
\Delta \chi^2=\sum_{i,j}( X_i- {\hat X}_i)
(\sigma^2)^{-1}_{ij}( X_j- {\hat X}_j)\, ,
\end{equation}
where $ {\hat X}_i$  are the central values of the electroweak parameters from the 
fit in Eq.~\ref{delts}, 
$ X_i$ are the contributions to these parameters from the new 
mirror fermions and from the Higgs loops,  and $\sigma^2_{ij}\equiv \sigma_i\rho_{ij}\sigma_j\,$, with $\, \sigma_i$ being the errors
given in Eq.~\ref{delts}.

We consider the case of no mass splitting within the doublets, while the 
fractional mass difference between the two heavy families is parametrized by $\delta$, 
and  focus on the regions of parameter space where we expect the  largest deviations with respect to 
the Standard Model $gg\rightarrow HH$ amplitude,  
while the single Higgs rate remains very close to the Standard Model value. 
Following the discussion in the previous section,
we therefore 
fix $\theta^t_-  = {\pi \over 2}$,  $\Delta = \{-0.1;\,0;\,0.1\}$ and choose $M =800$~GeV. 
%
\begin{figure}[tb]
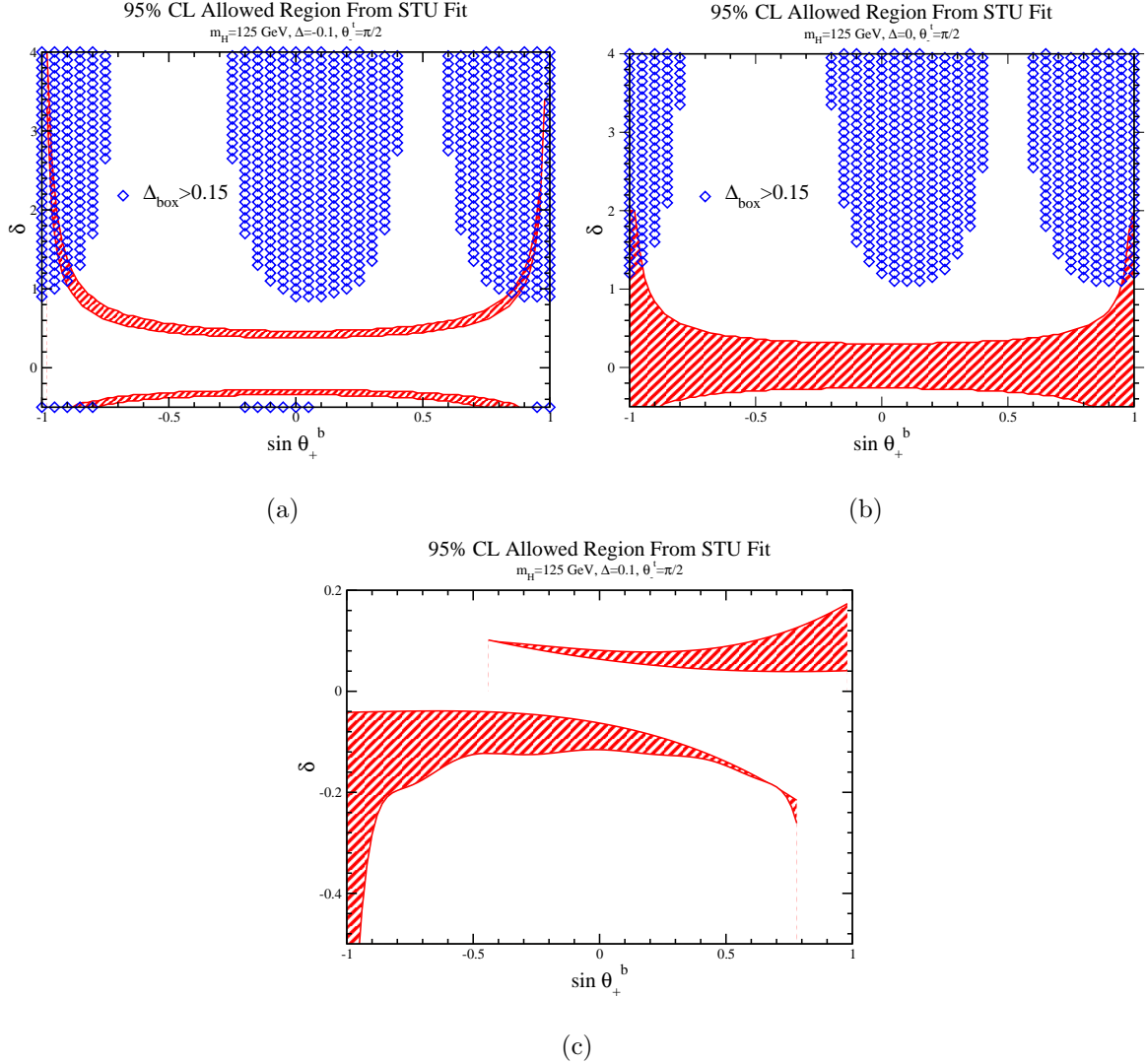

\subfigure[]{
      \includegraphics[width=0.45\textwidth,clip]{stu_d-1.eps}
      \label{fig:stu_d-1}
}
\subfigure[]{
      \includegraphics[width=0.46\textwidth,clip]{stu_d0.eps}
      \label{fig:stu_d0}
}
\subfigure[]{
      \includegraphics[width=0.46\textwidth,clip]{stu_d+1.eps}
      \label{fig:stu_d1}
}
\caption{Red bands: $95\%$ confidence level allowed regions  from the fit to electroweak data for single Higgs 
amplitudes which are suppressed~(a)/enhanced~(c) by $10\%$ with respect to the 
Standard Model amplitude, or equal to the Standard Model amplitude~(b),
for $\theta^t_- = {\pi \over 2}$ and $M =800$~GeV. Blue diamonds: parameter space regions which allow an 
enhancement of $15 \%$ or more to the double Higgs rate from the box topology. Such a large enhancement is 
not allowed by electroweak precision bounds in the case of $\Delta  = 0.1$~(c). }
\label{fig:ewpt_par_space}
\end{figure}
In Fig.~\ref{fig:ewpt_par_space} 
we show the $95\%$ allowed 
regions in the $\{ \sin \theta^b_+, \delta \}$ parameter space for the three values of $\Delta$ (red bands), 
along with the regions where the box enhancement is larger than $15\%$ (blue diamonds). 
The experimental bounds typically require $\delta$ to be small. In this limit, the electroweak parameters 
assume simple expressions,
	\bea
	\Delta S_F &=& {N_C \over 30 \pi} \Delta \left[ \frac{25}{6} + 4 \delta \sin \theta^b_+ 
		- \Delta \left( 1 + \delta \sin \theta^b_+ \right) +{\cal O}(\delta^2)\right] \; , \nn
	\Delta T_F &=& {N_C   \over 96 \pi  s_W^2 } {M^2 \over M_W^2} \Delta^2
		\left[ 2  + \delta \left(\sin \theta^b_+ + 2 \right) +{\cal O}(\delta^2) \right] \; , \nn
	\Delta U_F &=& {N_C  \over 60 \pi  } \Delta^2 \left[ 2 + \delta \sin \theta^b_+ 
	+{\cal O}(\delta^2) \right] \; ,
\label{slims}
\eea
 where $N_C=3$.
 For $\delta \to 0$, $\theta^b_- \to \theta^t_- = \frac{\pi}{2}$ and $\Delta \to 0$~(Eq.~\ref{SingHiggsLim.EQ}). 
 However, a large increase in the double Higgs  rate from the box topology can be obtained 
 only for large values of $\delta$. In particular, for $\Delta = 0.1$ the electroweak precision observables 
 do not allow the mass splitting to be large enough to obtain a significant enhancement, consistently 
 with the results from Fig.~\ref{fg:TwoHiggsAmp_Delt_pm01_0}.

\subsubsection{Phenomenology of the Mirror Fermion Model and $H \to \gamma \gamma$}

Once the parameters of the model are constrained to reproduce the Standard Model single Higgs amplitude
to within $\pm 10\%$ and to be allowed by a fit to the precision electroweak data, there is very
little freedom left to adjust parameters. 
The differential  cross section 
for $gg\rightarrow HH$ is shown for allowed parameters in 
\begin{figure}[tb]
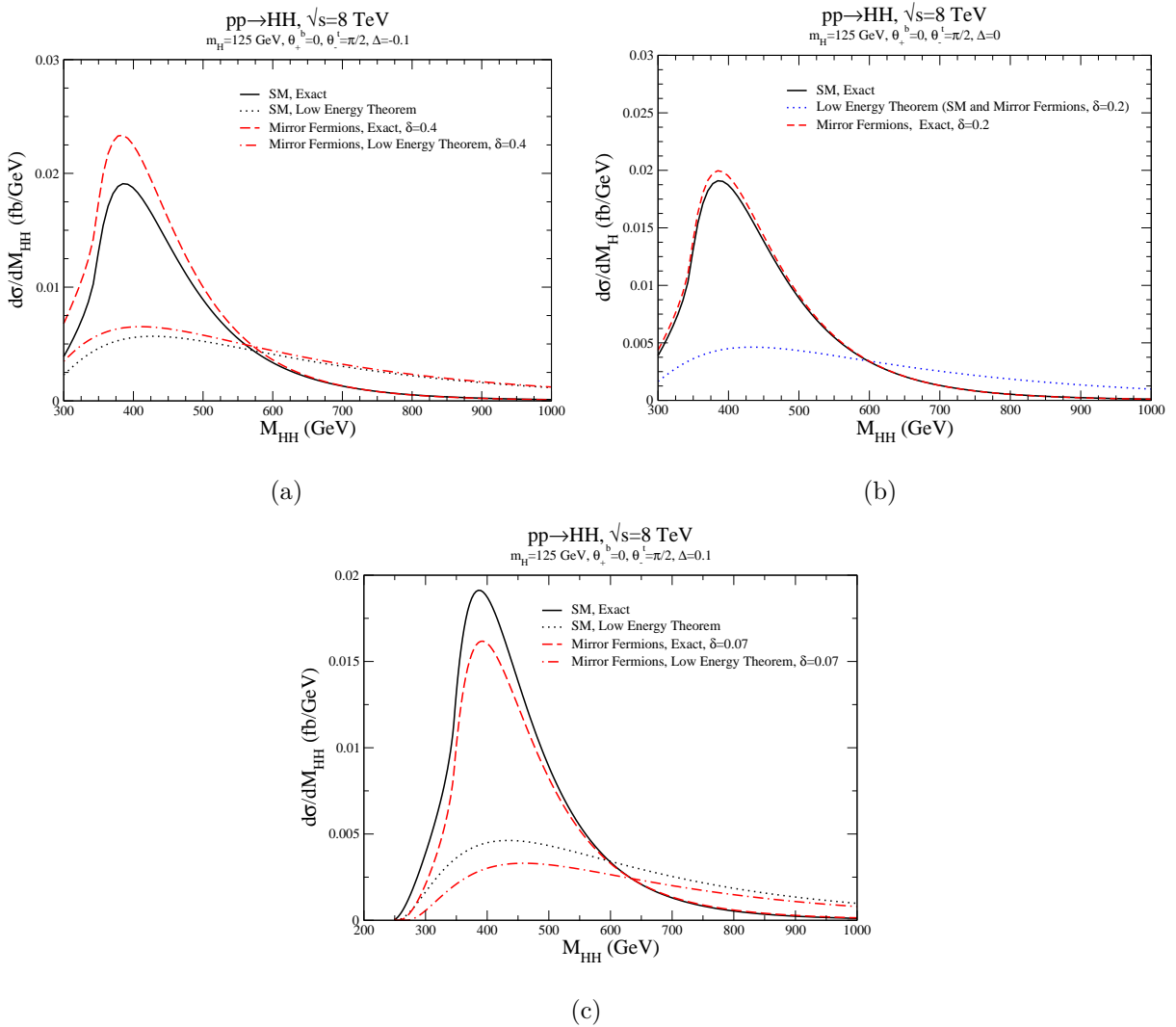

\subfigure[]{
      \includegraphics[width=0.47\textwidth,clip]{dsig_d-1.eps}
      \label{fig:dsig_d-1}
}
\subfigure[]{
      \includegraphics[width=0.48\textwidth,clip]{dsig_d0.eps}
      \label{fig:dsig_d0}
}
\subfigure[]{
	      \includegraphics[width=0.48\textwidth,clip]{dsig_d+1.eps}
      \label{fig:dsig_d1}
}
\caption{Differential double Higgs production cross section in the Standard Model and in the
mirror fermion model 
for $\theta^t_- = {\pi \over 2}$, $M =800$~GeV. The single Higgs production 
amplitude with respect to the Standard Model is suppressed by $10\%$ (a), equal (b) or enhanced by 
$10\%$ (c). We use CT10NLO PDFs and $\mu=M_{HH}=\sqrt{s}$. The curves labelled Low Energy 
Theorem use the infinite mass approximation to the rate.}
\label{fig:dsig_mirror}
\end{figure}
%
Fig.~\ref{fig:dsig_mirror} and it is clear that this class of
models does not allow for a large enhancement of the $HH$ production rate. 
The exact
cross sections include both Standard Model $t$ and $b$ contributions, while the
low energy theorem curves include the infinite mass limit of the heavy quark contribution.
The largest allowed enhancement is found for $\Delta=-0.1$ and in this case, the total cross 
section, $pp\rightarrow HH$ is enhanced by $\sim 17\%$ over the Standard Model rate.

The mirror fermions also contribute to the rate for $H\rightarrow \gamma\gamma$\footnote{We consider
only the contributions of heavy mirror quarks.  Heavy
leptons can also affect the $H\rightarrow \gamma\gamma$ rate~\cite{Carena:2012xa,Joglekar:2012vc,ArkaniHamed:2012kq,Kearney:2012zi,
Almeida:2012bq}.}.
We again consider each mirror family to be degenerate between the charge ${2\over 3}$
and charge $-{1\over 3}$ quarks, and the two families to be split 
by a mass difference $M \delta$. 
In the limit 
$m_H<<m_t,M_W, M$~\cite{Gunion:1989we}, 
\bea
	\sqrt{{\Gamma(H\rightarrow \gamma \gamma)
	\over\Gamma(H\rightarrow\gamma\gamma)_{SM}}}
	&=&
	1-{16\over 47}\biggl[
	{c_{T_1T_1}\over 2 M_{T_1}}
	+{c_{T_2T_2}\over 2 M_{T_2}}+
	{1\over 4}\biggl({c_{B_1B_1}\over 2 M_{B_1}}+{c_{B_1B_1}\over 2 M_{B_2}}\biggr)\biggr]
	\nonumber \\
	&=&
	1 - \frac{8}{47} \left[ 5 + \sin \theta^b_- \left( 3 \sin \theta^b_- - 8 \sin \theta^t_-\right)\right]
	\nn & & \quad
		-\frac{32}{47} \delta \sin \theta^b_+ \left( \sin \theta^b_- - \sin \theta^t_-\right)
		+{\cal O}(\delta^2) \;,
\eea
where we  impose only the angle relation from Eq.~\ref{eq:special_reln_masses_angles_PM} 
and expand for small $\delta$. In the limit $\delta = 0$ (and therefore $\theta^b_- = 
\theta^t_-$ from Eq.~\ref{eq:special_reln_masses_angles_PM}), the branching ratio 
into photons cannot be larger than in the Standard Model. 

We relate the deviations in the photon decay branching 
ratio to the deviation $\Delta$ from the Standard Model single Higgs production rate\footnote{This
result holds for arbitrary values of the parameters.}, 
\bea
	\sqrt{{\Gamma(H\rightarrow \gamma \gamma)
	\over\Gamma(H\rightarrow\gamma\gamma)_{SM}}}
	&=&
	1 - \frac{24}{47} \left[ \frac{ 4 (\delta+1)}{ ((\delta+2) \sin \theta^t_- -\delta \sin \theta^b_+ )^2}
	- \frac{(\delta+2) \sin \theta^t_- + \delta \sin \theta^b_+ }
		{(\delta+2) \sin \theta^t_- - \delta \sin \theta^b_+  }
		\right]
\nn & &
	- \frac{4}{47} \Delta \left[ 1 + \frac{3 (\delta+2) \sin \theta^t_- }
		{(\delta+2) \sin \theta^t_- - \delta \sin \theta^b_+} 
	-\frac{12 (\delta+1) }{((\delta+2) \sin \theta^t_- - \delta \sin \theta^b_+ )^2} 
	 \right] \nn
&& 
	- \frac{6}{47} \Delta^2 \frac{\delta+1}{((\delta+2) \sin \theta^t_- - \delta \sin \theta^b_+ )^2}  \;.
\eea

Imposing only the bounds from electroweak precision observables, and performing 
a general scan over the input parameters $\delta, \theta^b_+, \theta^b_-, 
\theta^t_+$ (fixing  $\theta^t_-$ through Eq.~\ref{eq:special_reln_masses_angles_PM}, 
$M=800$~GeV and $\delta$ in the range $\{-0.5; 2 \}$), we find that the Higgs branching ratio 
into photons can have large differences
 from the Standard Model predictions, with suppressions as large as $90\%$ 
and enhancements up to $10\%$. Requiring also the single Higgs production 
rate to be close to the Standard Model value puts severe constraints on these deviations. 
For a single Higgs production amplitude equal 
to the Standard Model prediction, the maximum deviation in the Higgs branching 
ratio into photons is $\pm 5\%$. 
For the regions of parameter space of Fig.~\ref{fig:stu_d-1}, 
where  $\theta^t_- = \frac{\pi}{2}$ and a $- 10\%$ deviation from the Standard Model prediction
 for the $gg\rightarrow H$ rate is allowed, only 
small enhancements (up to $+10\%$) of the $H\rightarrow\gamma\gamma$
rate are allowed. 
For a $+10\%$ enhancement in the single Higgs rate  over the Standard Model prediction 
(Fig.~\ref{fig:stu_d1}), the branching ratio into 
photons deviates from the Standard Model prediction by at most by a few percent. 
\begin{figure}[tb]
      \includegraphics[width=0.6\textwidth]{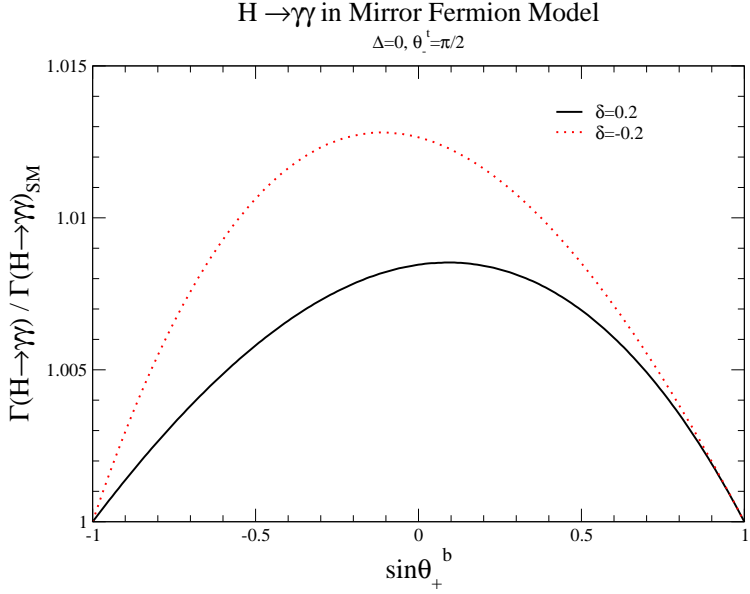}
\caption{Ratio of $\Gamma(H\rightarrow \gamma\gamma)$
to the Standard Model Higgs branching ratio into photons 
for the points of Fig.~\ref{fig:stu_d0}, 
where $\Delta=0$ and $\theta_-^t={\pi\over 2}$. We fix 
$\delta = \pm 0.2$, which is allowed from the electroweak fit for all the 
values of $\theta^b_+$. Larger deviations from the Standard Model 
$H \to \gamma \gamma$ branching ratios arise outside this range of 
$\delta$, in the regions where $|\sin \theta^b_+|$ is close to 1.}
\label{diags:Rgamma}
\end{figure}
We show how these deviations depend on the free input parameters 
$\delta, \sin \theta^b_+$ in Fig.~\ref{diags:Rgamma}, where we focus 
on $\Delta = 0$ and pick two values of $\delta$ which are allowed by 
the electroweak fit over all the range of $\theta^b_+$ (with $\theta_-^t={\pi\over 2}$). 
The clear conclusion is that the restrictions from precision electroweak data, combined
with a single Higgs production rate close to the Standard Model
prediction, do not allow for significant deviations in the $H\rightarrow \gamma
\gamma$ rate in this class of models.

\section{Connection to gluon-Higgs dimension six operators}
\label{sec5int_dim6}
An interesting idea~\cite{Pierce:2006dh} is to combine single and double Higgs production 
to gain insights on the mechanism giving mass to the particles that contribute to these 
loop-mediated processes. Including contributions up to dimension-6 operators, 
the effective Lagrangian responsible for the Higgs-gluon
interactions can be written as
\begin{equation}
{\cal L}=c_1{\cal O}_1 +c_2 {\cal O}_2\; .
\end{equation}
Particles whose mass arises entirely from renormalizable Higgs couplings induce an operator 
\beq
{\cal O}_2 
= 
{\alpha_s \over 24\pi } G^a_{\mu \nu} G^{a,\mu \nu} 
\log \left( {\Phi^\dagger \Phi \over v^2} \right) 
\simeq
{\alpha_s \over 12 \pi } G^a_{\mu \nu} G^{a, \mu \nu}
\left({H \over v} - {H^2 \over 2 v^2} \right)
\; .
\eeq
If the particle receives contributions to its mass from other sources as well, 
an additional operator 
\beq
{\cal O}_1 
= 
{\alpha_s \over 12 \pi } G^a_{\mu \nu} G^{a,\mu \nu} 
 \left( {\Phi^\dagger \Phi \over v^2} \right) 
 \simeq
  {\alpha_s \over 12 \pi } G^a_{\mu \nu} G^{a,\mu \nu} 
 \left({H \over v} + {H^2 \over 2 v^2} \right) 
\eeq
arises. In the Standard Model $c_1^{SM} = 0,$ $c_2^{SM} = 1$. 
The two operators contribute differently to Higgs single and pair production and 
the different rates in these channels constrain the 
coefficients $c_1$ and $c_2$. 
Following~\cite{Pierce:2006dh}, one can derive these two coefficients in a 
background field approach. The Higgs field is treated as a background field, 
and the masses of the heavy particles become thresholds in the running of $\alpha_s$. 
Matching the low- and high-energy theories~\cite{Shifman:1979eb,Kniehl:1995tn}, 
\beq
	\frac{1}{g_{eff}^2(\mu)} = \frac{1}{g_s^2(\mu)} - \frac{\delta b_f}{8 \pi^2} \log \det \frac{{\cal M}(H)}{\mu} \;,
\eeq
where ${\cal M}(H)$ is the Higgs-dependent mass matrix and  $\delta b_f=2/3$ 
for fermions in the fundamental representation of the color group.
This yields the effective Lagrangian
\beq
	{\cal L}_{eff} = \frac{\alpha_s}{12 \pi} G^a_{\mu \nu} G^{a,\mu \nu} 
		\log \det {\cal M}(H) \; .
\label{Leff.EQ}
\eeq
We write the determinant of the mass matrix as 
\beq
\det{\cal M}(H)=\left[1+F_i(H/v)\right]\times P(\lambda_i,m_i,v) \; ,
\label{fact.EQ}
\eeq
where $P$ is a polynomial of the Yukawa couplings $\lambda_i$ and fermionic masses $m_i$ 
and in general $F_i(H/v) = F(H/v,\lambda_i,m_i,v)$. 
If $F_i(H/v)$ is such that 
$ F'_i(0)=1 + F_i(0)$, and all the higher order derivatives vanish before electroweak symmetry breaking, 
then the Higgs production rates via gluon fusion in the heavy quark limit are exactly as in the 
Standard Model\footnote{
For the purpose of this discussion, we only need $F_i''(0) = 0$. Nonvanishing derivatives at higher 
orders only affect gluon fusion production of three or more Higgs bosons.}. 
This is the case in the singlet top partner model, where $F_i(H/v) = H/v$ and therefore 
$c_{1,2} = c_{1,2}^{SM}$. 


Interestingly, one can have the same single Higgs production rate as in the Standard Model, but a different 
double Higgs rate, only for $F''_i(0) \neq 0$. If also the first condition, $ F'_i(0)=1 + F_i(0)$, is not met, 
then the single Higgs rate is not Standard Model like. In such a case, we note that for $F_i$ independent of Yukawa couplings and fermionic masses, the Higgs rates do not depend on the details of the fermion sector~\cite{Gillioz:2012se} and  deviations can arise only from changes to the Higgs potential.
 If $F_i$ depends on the Yukawa couplings and fermionic masses, 
 the Higgs rates will  in general be related 
 to these parameters. 
Such a situation occurs for example in the mirror fermion model. 
In this case 
\beq
	\begin{array}{rclcrll}
		c_1^t &=& \frac{-2 \beta_t}{(1 - \beta_t)^2}  & , \; &
		c_1^b &=&  \frac{-2 \beta_b}{(1 - \beta_b)^2} \, ,\\ \\
		c_2^t &=& 
	 1 +  \frac{2}{(1 - \beta_t)^2} & , \; & 
	 c_2^b &=&   \frac{2}{(1 - \beta_b)^2}  \, .
\end{array}
\eeq
We define 
\beq
	\beta_t = \frac{\lambda_E \lambda_F}{\lambda_B \lambda_D v^2/2} \quad , \qquad
	\beta_b = \frac{\lambda_E \lambda_G}{\lambda_A \lambda_C v^2/2} \; .
	\label{eq:betas}
\eeq
In terms of the physical parameters, 
\bea
	\beta_q &=& 
	1 - \frac{ 4 (1+\delta)  }{(2 + \delta)^2 \cos^2 \theta^q_- - \delta^2 \cos^2 \theta^q_+} \; , 
	\qquad q = t, b \; .
	\label{eq:betaq_delta}
\eea

For $\beta_b \to 0$, $c_1^b$ and $c_2^b$ go to twice the Standard Model value. 
In this limit, the vector contributions to the fermion mass matrix vanish, and the masses 
come entirely from electroweak symmetry breaking. Since there are two quarks, an 
extra factor of two arises. 
In $c_2^t$ one clearly sees the $+1$ contribution coming from the Standard
Model top quark.  

The coefficients governing single and double Higgs production are then 
\bea
	c_H &\equiv& c_1 + c_2 = 1 + 2 \left[ \frac{1}{1- \beta_t} +  \frac{1}{1- \beta_b} \right] \; ,
	\nn
	c_{HH} &\equiv& c_1 - c_2 = - 1
			- 2\left[ \frac{1+ \beta_t}{(1- \beta_t)^2} +  \frac{1+ \beta_b}{(1- \beta_b)^2} \right] \; .
\label{eq:cHHH_mirror}
\eea
The two rates depend on the two independent parameters $\beta_t, \beta_b$ from the top 
and bottom sectors. 
Even if we require the single Higgs rate, $gg\rightarrow H$, to be close to the Standard Model value,
\beq 
c_H = c_1 + c_2 \rightarrow  c_H^{SM}(1+\Delta)=1+\Delta \; ,
\label{chcon}
\eeq
we are left with an independent parameter that can yield completely independent variations 
in the double Higgs rate.
 
 \begin{figure}[tb]
      \includegraphics[width=0.6\textwidth]{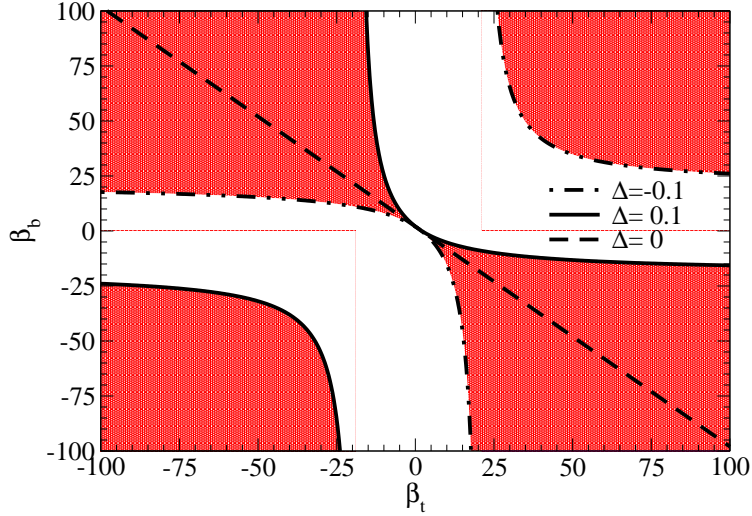}
\caption{The shaded red regions correspond to amplitudes for $gg\rightarrow H$ within
$\pm 10\%$ of the Standard Model rate.}
\label{fg:betas}
\end{figure}
In Fig.~\ref{fg:betas}, we show the regions of $\beta_b$ and $\beta_t$ which reproduce
the Standard Model Higgs amplitude to within $\Delta=\pm 10\%$. 
 Imposing the constraint of Eq.~\ref{chcon}  on the single Higgs rate in general constrains the double Higgs rate, $gg\rightarrow HH$,
 \bea
	c_{HH} \rightarrow 2c_1-(1+\Delta)\, .
\eea
In the singlet case, $c_1=0$ and deviations in single and double Higgs rates must be of the same 
order of magnitude. In the mirror case, $c_1$ can deviate from zero, which removes the close relationship 
between single and double Higgs production.

In terms of the parameters of the mirror fermion model,
\bea
	c_{HH} &\rightarrow & c^{SM}_{HH}\left(1+{8\over (1-\beta_t)^2}-{5-\beta_t\over 1-\beta_t}\Delta+\Delta^2\right)
	 \nn
	&=& - \left\{ 1+   \frac{1}{2} \left[ \frac{
		(2+\delta)^2 \cos^2 \theta^t_-  - \delta^2 \cos^2 \theta^t_+}{1+\delta}
	\right]^2 +{\cal O}(\Delta)\right\}  \;.
\eea
The term in the curly brackets correctly reproduces $1+\Delta_{box}$ from Eq.~\ref{eq:deltabox_smalldD} 
for $\Delta=0, \theta^t_- = \frac{\pi}{2}$. 
\begin{figure}[tb]
 \includegraphics[width=0.6\textwidth,clip]{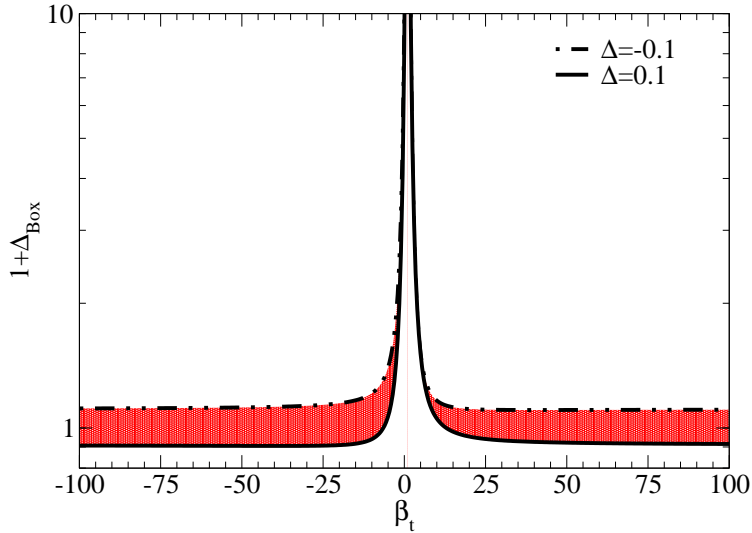}
\caption{Enhancement of the box contribution to $gg\rightarrow HH$ for a single Higgs amplitude 
within $\pm 10\%$ of the Standard Model prediction.}
\label{fg:betadouble}
\end{figure}
A large effect in the double Higgs rate requires large $c_1$, and, in turn, $\beta_t\sim 1$. 
This is seen in Fig.~\ref{fg:betadouble}, 
where we fix $\beta_b$ to reproduce the single Higgs rate within $10\%$ the Standard Model value.    
However, from Eq.~\ref{eq:betaq_delta} $\beta_t \rightarrow 1$ implies 
$\delta\rightarrow -1$ or $\delta \rightarrow \infty$. 
These are not viable solutions. The first one corresponds to massless quarks. The second one 
requires non-perturbative interactions with the Higgs (large $\lambda_B, \lambda_D)$ 
for heavy quarks (large $\lambda_E, \lambda_F)$, as in Eq.~\ref{eq:betas}. 
In the mirror fermion model discussed
 in this paper, large deviations in the $gg\rightarrow HH$ rate do not occur.

\section{Conclusions}
\label{sec5_conclusions}
We analysed double Higgs production from $gg\rightarrow HH$ in the Standard Model and in models with additional 
heavy vector or chiral quarks. In the Standard Model, we compared the approximate results 
in the large top mass expansion with the exact cross section, and analysed the dependence 
of the production rate on the choice of the renormalization/factorization scale $\mu$ 
and on the PDF sets. 
As is well known~\cite{Baur:2002rb,Binoth:2006ym}, 
the low energy theorems fail to accurately reproduce both the total and differential double Higgs 
cross sections. 
The differential distributions  are poorly estimated by the low energy theorems
and predict a large tail at high invariant masses. 
The discrepancy is smallest for 
the scale choice $\mu = 2 m_H$,  yielding a $10-25\%$ difference
from the exact calculation of the total rate.  
Further, the predictions of the large top mass expansion depend sensitively
on the choice of PDFs.
Inclusion of higher order terms in the large mass expansion does not improve
 the convergence towards the exact results.

We discussed how the combination of single and double Higgs production from gluon fusion
might give insight into the mechanism giving mass 
to  quarks.   The parameters of models with new heavy fermions are strongly constrained both
by the observed rate for $gg\rightarrow H$ and by precision electroweak measurements.  
In the case of a new heavy vector singlet quark, electroweak precision observables 
strongly constrain its mixing with the top 
quark~\cite{Dawson:2012di}. The singlet needs almost to decouple from the Standard
Model particles, and therefore deviations from the Standard 
Model in 
both the single and double Higgs rates
are small.

The situation is more interesting in the case of heavy mirror quarks which are not allowed
to mix with the Standard Model fermions. The bounds 
from electroweak precision data still allow for the single Higgs production cross section to   
differ from the Standard Model predictions. However, after restricting the deviations in the 
 $gg\rightarrow H$ rate from the Standard Model rate to be small,  the resulting 
 double Higgs cross section and distributions become close to those of the Standard Model. 
The reason for this behaviour becomes clear in terms of the two dimension six operators 
${\cal O}_1$ and ${\cal O}_2$. Once we fix the single Higgs rate to be close to 
that of the Standard Model, 
large deviations in the double Higgs rate occur only if one of the mirror family becomes very 
heavy, with non-perturbative Higgs interactions, or very light, outside the range $m_H < 2 m_q$ where 
the operator expansion applies.
In the mirror fermion model we also investigated the effects of the additional quarks on 
the Higgs branching 
ratio to photons.  After the constraints from the observed single Higgs 
cross section and precision electroweak measurements
 are taken into account, the branching ratio $H\rightarrow \gamma \gamma$
 is always within
  $10\%$ 
  of the Standard Model rate.
  
  Therefore, in the two example of models with new heavy fermions which we
  studied, the constraints 
  from the observed $gg\rightarrow H$ rate, combined with precision electroweak
  data,  do not allow large deviations of the $gg\rightarrow HH$ rate from the Standard Model
  prediction.

\section*{Acknowledgements}
This work is supported by the U.S. Department of Energy under grant
No.~DE-AC02-98CH10886. 

\appendix
\section{Electroweak parameters in the mirror fermion model} 
\label{app1_ew_mirror}
We present here some useful formulae for the mirror fermion model. 

The parameters $\lambda_i$ appearing in the mass Lagrangian~\ref{eq:lag_mirror} 
can be expressed in terms of the physical masses and mixing angles as 
\bea
	\lambda_2 {v\over \sqrt{2}} & = & m_t \; , \nn
	\lambda_B {v\over \sqrt{2}} & = & M_{T_1} \cos \theta^t_L \cos \theta^t_R + M_{T_2} \sin \theta^t_L \sin \theta^t_R \; , \nn
	\lambda_D{v\over \sqrt{2}} & = & M_{T_1} \sin \theta^t_L \sin \theta^t_R + M_{T2} \cos \theta^t_L \cos \theta^t_R \; , \nn
	\lambda_E &=& M_{T_2} \sin \theta^t_L \cos \theta^t_R - M_{T_1} \cos \theta^t_L \sin \theta^t_R \; , \nn
	\lambda_F &=& M_{T_2} \cos \theta^t_L \sin \theta^t_R -M_{T_1} \cos \theta^t_R \sin \theta^t_L \; .
\eea
Similar relations hold for the corresponding parameters in the bottom sector, with $M_{T_i} \to M_{B_i}$ and 
$\theta^t_{P} \to \theta^b_{P}$.

The charged current interactions among quarks of charge $Q$ and $(Q-1)$ are 
\bea
\label{eq:LCCM}
	\lag^{CC}_M 
	&=& 
	\frac{g}{\sqrt{2}} \sum_{i,j} \left\{
		\bar{q}^i_{Q} \gamma^\mu
		\left[
			V^{L}_{ij} P_L+V^{R}_{ij} P_R 
		\right]
		q^j_{(Q-1)} 
	\right\}
	W^+_\mu + \mathrm{h.c.}\,,
\eea
with
\beq
\begin{array}{rclccrll}
	V^L_{T_1B_1} &=& \cos \theta_L^b		\cos \theta_L^t &\quad &
	V^L_{T_1B_2} &=& \sin \theta_L^b		\cos \theta_L^t & \\	
	V^L_{T_2B_1} &=& \cos \theta_L^b 	\sin \theta_L^t & \quad &
	V^L_{T_2B_2} &=& \sin \theta_L^b 	\sin \theta_L^t & \\	
	V^R_{T_1B_1} &=& \sin \theta_R^b		\sin \theta_R^t & \quad &
	V^R_{T_1B_2} &=& -\cos \theta_R^b		\sin \theta_R^t &  \\	
	V^R_{T_2B_1} &=& -\sin \theta_R^b 	\cos \theta_R^t & \quad &
	V^R_{T_2B_2} &=& \cos \theta_R^b 	\cos \theta_R^t  \, .&
\end{array}
\eeq
We can rewrite these relations as 
\bea
	V^L_{ij} &=& \left( U_L^t \right)_{i1} \left( U_L^b \right)_{j1} \, , \qquad
	V^R_{ij} = \left( U_R^t \right)_{i2} \left( U_R^b \right)_{j2} \,.
\eea

The neutral current interactions among quarks of charge $Q$ are 
\bea 
\label{eq:LNCM}
	{\cal L}^{NC}_M
	&=&
	\frac{g}{2c_W} \sum_{i,j} \left\{
		\bar{q}^i_{Q} \gamma^\mu 
		\left[ 
			X^{L}_{ij} P_L +X^{R}_{ij} P_R
			-2 s_W^2 Q \delta_{ij} \right] q^j_{Q} 
	\right\}
	Z_\mu + {\rm h.c.} \, ,
\eea
where
\beq
\begin{array}{rclcrclcrclr}
	X^L_{T_1 T_1} &=& \cos^2 \theta_L^t & \quad &
	X^L_{T_1 T_2} &=& X^L_{T_2 T_1}  = \phantom{-}\sin \theta_L^t	 \cos \theta_L^t &  \quad &
	X^L_{T_2 T_2} &=& \sin^2 \theta_L^t & \\	
	X^R_{T_1 T_1} &=& \sin^2 \theta_R^t & \quad &
	X^R_{T_1 T_2} &=& X^R_{T_2 T_1}  = - \sin \theta_R^t	\cos \theta_R^t &  \quad &
	X^R_{T_2 T_2} &=& \cos ^2\theta_R^t &\;.
\end{array}
\eeq
The same relations, up to an overall minus sign, hold in the bottom sector. In more compact 
form we can write 
\beq
	X^L_{ij} = \pm \left( U_L^{t,b} \right)_{i1} \left( U_L^{t,b} \right)_{j1} \, , \qquad
	X^R_{ij} = \pm \left( U_R^{t,b} \right)_{i2} \left( U_R^{t,b} \right)_{j2} \, ,
\eeq
where the plus sign holds in the top sector and the minus in the bottom sector. 

\bibliographystyle{hunsrt}
\bibliography{paper}

\end{document}